\documentclass[aps,preprint]{revtex4}
\usepackage{graphicx}
\usepackage{epsf}
\usepackage{wrapfig}
\usepackage{epsfig}

\def\b0{{\mbox{\boldmath$0$}}}
\def\dfrac{\displaystyle\frac}

\def\b0{{\mbox{\boldmath$0$}}}

\newcommand{\ra}{\,\rangle}
\newcommand{\la}{\,\langle}

\def \b #1{ {\bf #1}}
\newcommand{\be}{\begin{eqnarray}}
\newcommand{\ee}{\end{eqnarray}}
\def\Spp{^3S_1^{++}}
\def\Dpp{^3D_1^{++}}

\def \b #1{ {\bf #1}}

\def\Spp{^3S_1^{++}}
\def\Dpp{^3D_1^{++}}

\newcommand{\CM}{{\cal M}}

\def\dfrac{\displaystyle\frac}

\def \b #1{ {\bf #1}}

     \font\tenbifull=cmmib10 scaled 1200 
     \font\tenbimed=cmmib9
     \font\tenbismall=cmmib7
       \def\bmit{\fam9 }
\mathchardef\bbkappa="7114
\mathchardef\bbrho="711A
\mathchardef\bbsigma="711B
\mathchardef\bbtau="711C
\mathchardef\bbvarrho="7125
\mathchardef\bbvarsigma="7126
\mathchardef\bbxi="7118
\def\boldkappa{{\bmit\bbkappa}}
\def\boldrho{{\bmit\bbrho}}
\def\boldsigma{{\bmit\bbsigma}}

\def\boldxi{{\bmit\bbxi}}
\begin{document}
\vskip 2mm \date{\today}\vskip 2mm
\title{ Calculations of the  Exclusive Processes  $\bf ^2H(e,e'p)n$,
  $\bf ^3He(e,e'p)^2H$  and  $\bf ^3He(e,e'p)(pn)$   within a Generalized Eikonal Approximation}
\author{C.Ciofi degli Atti}
\author{L.P. Kaptari}
\altaffiliation{On leave from  Bogoliubov Lab.
      Theor. Phys.,141980, JINR,  Dubna, Russia}
\address{Department of Physics, University of Perugia and
      Istituto Nazionale di Fisica Nucleare, Sezione di Perugia,
      Via A. Pascoli, I-06123, Italy}
 \vskip 2mm

\begin{abstract}
\vskip 5mm
The exclusive  processes $^2H(e,e^\prime p)n$,
$^3He(e,e^\prime p)^2H$ and  $^3He(e,e'p)(pn)$,
have been analyzed using realistic few-body wave functions and treating
the final state interaction (FSI)  within a Generalized Eikonal Approximation (GEA), based
upon  the direct calculation of the  Feynman diagrams describing  the rescattering
of the struck nucleon with the nucleons of the $A-1$ system.
The approach  represents
an improvement of the conventional Glauber    approach (GA), in that
it allows one to
take into account the effects of the nuclear excitation
of the  $A-1$ system
on the rescattering
of the struck nucleon.
Using realistic three-body wave functions corresponding to the $AV18$ interaction, the results
of
 our parameter free  calculations  are compared with available experimental data.
  It is found that in some kinematical conditions FSI effects
 represent small corrections, whereas in
other kinematics conditions
they are very large and absolutely necessary  to provide a
satisfactory agreement between theoretical  calculations and experimental data.
 It is shown that in the kinematics of the
experimental data which have been considered,
 covering the  region of missing momentum and energy with
 $ p_m \leq\, 0.6 \, \, GeV/c$ and  $E_m\,
 \leq 100 \,MeV$
in the perpendicular kinematics,
the GA and GEA predictions  differ
only by less than  $\simeq 3-4\%$.

\end{abstract}
\maketitle

 \section{Introduction}
One of the main aims of  nowadays hadronic physics is the investigation of the
limits  of validity  of the so called \textit{Standard Model} of nuclei, i.e.
the description of nuclei in terms of the solution of the non relativistic
Schr\"odinger equation containing realistic nucleon-nucleon interactions.
To this end, exclusive lepton scattering could be very useful, for it might
yield relevant information on the nuclear wave function,
provided  the initial and final states involved in the scattering process
 are described within a consistent, reliable approach. In the case of
\textit{few-body systems},  a consistent treatment of initial and final
states is nowadays possible at low energies (see e.g. \cite{gloeckle,pisa}
and References therein quoted), but at higher energies,  when the number of
partial waves sharply increases  and  nucleon-nucleon (NN) interaction becomes highly
inelastic,  the
Schr\"odinger approach  becomes impractical and other methods have to be employed.
In the case of \textit{complex nuclei}, additional difficulties arise
 due to the approximations which are still necessary to solve
the many-body problem. As a matter of fact, whereas  fundamental progress
has been made in recent years in the calculation of various properties
 of light nuclei
(see e.g. \cite{wiri} and  \cite{pie01} and References therein quoted),
 much remains  to be done for the treatment of the continuum,
 for which various approximate treatments of the final state  cannot be avoided.
 In this context, it should be
 stressed that calculations involving  few-body systems,  where the ground state can be treated
exactly, can also be very useful to investigate the limits
of validity of various approximate schemes to treat the continuum and
their possible extension to complex nuclei.

The aim of this paper is to present the results of a systematic  theoretical investigation
of the exclusive process $A(e,e'p)B$  off $^2H$ (to be also denoted by $D$) and $^3He$,
based on a reliable description  of:
\begin{enumerate}
\item  initial state
correlations (ISC), treated by the use of  the status-of-the-art
few-body wave functions~\cite{pisa} corresponding to the $AV18$
interaction \cite{av18};
\item final state interactions (FSI),
treated within a relativistic
 framework based upon the calculation of the relevant
  Feynman diagrams which describe  the rescattering of the
   struck nucleon
  by the other $A-1$  \textit{spectator} nucleons of the target.
\end{enumerate}

 Whereas a correct  treatment of ISC in few-body systems is automatically achieved
 by the use of realistic wave functions, the treatment of FSI at high energies
 is still matter of
 discussions. The approach we are going to use has several non trivial advantages,
 in that it allows one to work within a relativistic framework provided by
 the use of Feynman diagrams and, moreover, it can be applied, in principle,
 to the treatment of exclusive $A(e,e'p)B$ processes off complex nuclei as well.
  It should be stressed, at this point,
  that the diagrammatic approach we are talking about
   is not a new one:
it has been first formulated
in Ref. \cite{gribov} and \cite{bertoc} (see also Ref. \cite{weis}),
within a spin-less treatment of particle-nucleus scattering,
  and applied subsequently to various
   types of high energy processes with nuclear targets.
   More recently,
 the diagrammatic approach  has been generalized
  to the treatment of the FSI in  exclusive
$A(e,e'p)B$  \cite{mark,strikman,misak} and
$A(e,e'2p)B$ \cite{marknew} processes, and a Feynman diagram
approach has also been used in  Ref. \cite{bra01,misha} and
\cite{hiko}  to take into account off-shell effects both in
inclusive, $A(e,e')X$, and exclusive,
 $A(e,e'p)B$, processes. 

The diagrammatic approach  we are referring to, is a generalization
of the standard  Glauber  Approach (GA)  \cite{glauber} based on the
eikonal approximation, so that, following Refs. \cite{marknew}, we
will call it
 Generalized Eikonal  Approximation (GEA).

  It is
 well known that  the application
of the GA to the treatment of $A(e,e'p)B$ processes requires the following
 approximations:
 i) the NN scattering amplitude is
 obtained within the eikonal approximation; ii) the
nucleons of the spectator  system  $A-1$ are stationary during the multiple scattering
with  the struck
nucleon  (the
{\it frozen approximation})  , and iii)
only perpendicular
 momentum transfer components in the NN scattering amplitude
are considered.
 In the GEA the    {\it frozen approximation} is partly removed by taking
 into account the excitation energy of the  $A-1$ system,
  which results in a correction term to the standard
 profile function of GA,  leading   to an additional contribution to the
  longitudinal component of the missing momentum.

In the present paper we apply both the GA and the GEA
 to the calculation of the processes
$^2H(e,e'p)n$, $^3He(e,e'p)^2H$, and  $^3He(e,e'p)(np)$,
 and compare our results with available
experimental data
       ~\cite{nikhef,ulmer,bulten, saclay,jlab1,jlab2}. The $^3He$ wave function  of the
   Pisa group \cite{pisa}, corresponding to the AV18 interaction \cite{av18}, will be used
   in the calculations.
   We  will not consider, for the time being,
   Meson Exchange Currents (MEC), $\Delta$-Isobar Configurations, and similar effects,
    which have been the object of intensive  theoretical studies in $A(e,e'p)B$
    processes off both few-body systems (see e.g. \cite{laget5,vanleuwe})
     and complex nuclei (see e.g. \cite{ryckebusch} and References therein quoted).
     As  in Refs. \cite{mark,strikman, misak,marknew}, we
   fully concentrate on the effects of the FSI but  we
    will consider kinematical conditions for which the effects
   from meson exchange currents (MEC) and $\Delta$
 excitation effects are  expected to be small corrections, and, whenever
 possible,  we will compare our results with the results by  other
 authors  which include these effects.

The structure of the paper is as follows:
in Section~\ref{sec:2} the basic formalism  of lepton-hadron scattering
is briefly illustrated and the
main formulae are obtained;
in Section~\ref{susecA} the concept of
Plane Wave Impulse Approximation (PWIA)  and  Spectral Function are recalled;
 in Section IV, the  GEA  is introduced, the relevant  Feynman diagrams  which one needs
to take into account in the treatment of the full FSI
are analyzed, and the problem of the factorization of the
lepton-nucleus  cross section within the GA and GEA   is also discussed; the
 results of the calculations and their comparison with available experimental data
are shown in Section~\ref{serR}; eventually, the Summary and
Conclusions  are presented  in Section~\ref{sec:4}. Some details
concerning  the formal aspects of our approach are given in Appendices A and~B.
Preliminary results of our calculations have
 been reported in Ref. \cite{greno} and \cite{pavia}.

\section{Basic formulae of $(e,e'p)$ scattering off nuclei}
 \label{sec:2}
 
\begin{figure}[!htp]
\hspace*{-2mm}                     
\epsfig{file=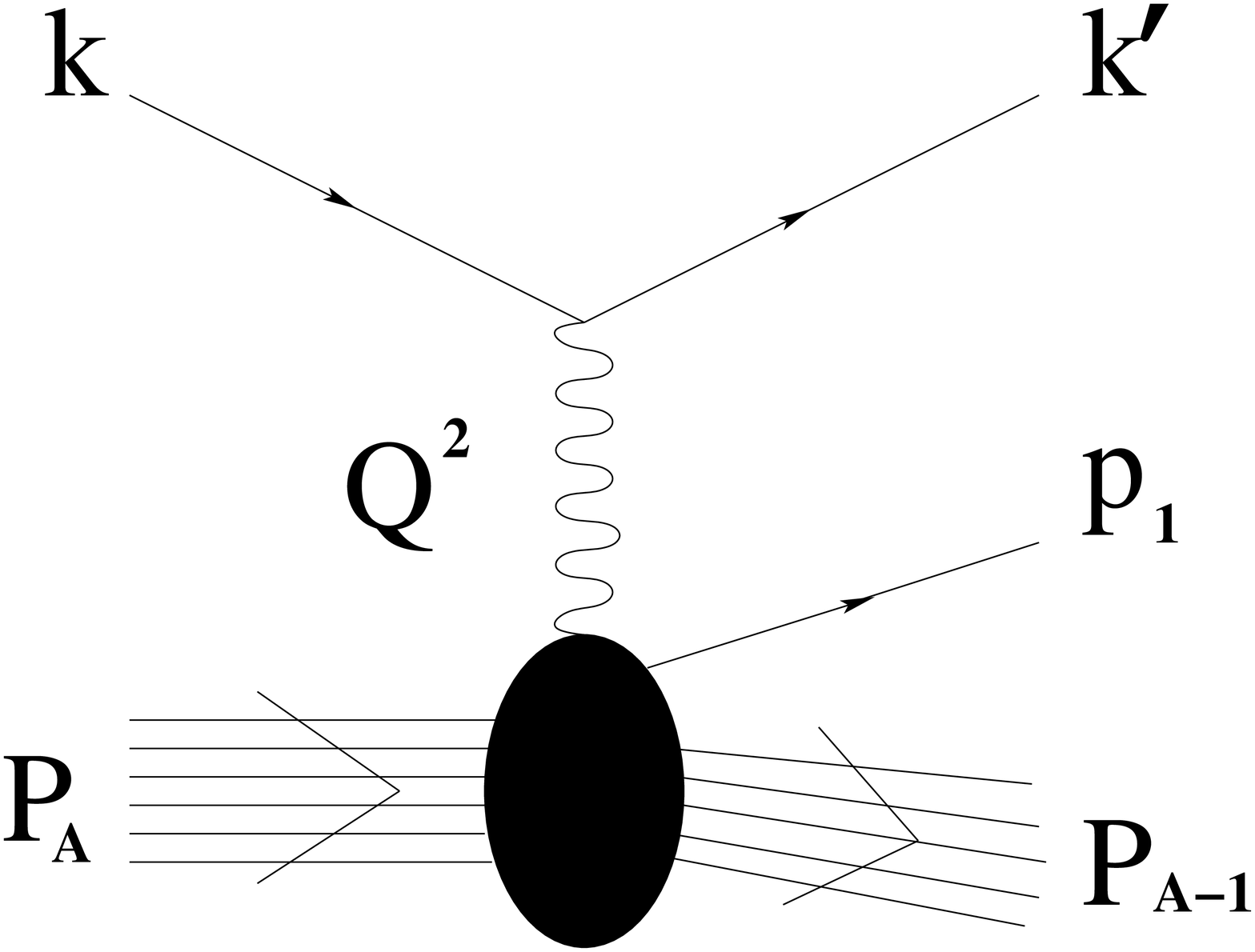,width=10cm,height=6.5cm}
\caption{ The  one-photon exchange approximation
for the process $A(e,e^\prime p)(A-1)$.}
      \label{fig1}
\end{figure}

The one-photon-exchange diagram for the process $A(e,e'p)(A-1)$, where $A-1$
 denotes a system of $A-1$ nucleons in a
bound or continuum state,  is presented in Fig. \ref{fig1},
where the relevant four-momenta in the scattering processes are shown, namely
the electron momenta before and after
 interaction,
 $k=(E,\b{k})$  and
 $k^{'}=(E^{'},{\b{k}}^{'})$,
  the momentum of the target nucleus  $P_A=(E_A,{\b{P}}_A)$  and, eventually,
   the momenta
  of the final
 proton and the final
 $A-1$ system,  $p_1=(\sqrt{{{\b{p}}_1}^2 +M_N^2},
 {{\b{p}}_1})$ and $P_{A-1}=(\sqrt{\b{P}_{A-1}^2
  +(M_{A-1}^{f})^2},{\b{P}}_{A-1})$,
   where $M_N$ is the nucleon mass,
   $M_{A-1}^{f}=M_{A-1}+E_{A-1}^f$,
and  $E_{A-1}^f$ is the {\it intrinsic} excitation energy of the $A-1$ system.

Let us briefly recall some useful formulae regarding the process
described  by the diagram shown in Fig. \ref{fig1}. The differential cross section
for the exclusive process  has the  following form (see e.g.  ~\cite{electron})
\be \!\!\!\!\raggedleft
\frac{d^6\sigma}{d E' d\Omega'\ d^3{{\b p}_1}}=
\sigma_{Mott}\tilde{{l}}^{\mu\nu}W_{\mu\nu}^A,
\label{eq1}
\ee

\noindent where
$\sigma_{Mott}=\displaystyle\frac{4\alpha^2\ {E'}^ 2\cos^2\frac\theta2}{Q^4}$
 is  the Mott cross section, $\alpha$  the fine-structure constant,
 $Q^2= -q^2 = -(k-k')^2 = {\b q}^2 - q_0^2 = 4EE'\sin^2\theta/2$   the four-momentum transfer,
 $ \theta \equiv
\theta_{\widehat{\b k \b k^{'}}}$ the scattering angle.
 The quantities $\tilde{{l}}_{\mu\nu} $ and $W_{\mu\nu}^A$ are the reduced
 leptonic
 and hadronic  tensors, respectively;  the former
 has the well known standard form (\cite{electron}), whereas the latter
 can be written as follows
\begin{eqnarray}
W_{\mu\nu}^{A} & = & \frac{ 1}{4\pi M_A} {\overline
{\sum_{\alpha_A}}} \sum_{ \alpha_{A-1}, \alpha_N}
(2\pi)^4 \delta^{(4)} (P_A + q - P_{A-1} -p_1)\times
\nonumber\\
 &\times&\langle \alpha_A \b P_A| {\hat J_\mu^A(0)} |
\alpha_N{{\b p}_1} , \alpha_{A-1}{\b P}_{A-1}  E_{A-1}^f \rangle
\langle  E_{A-1}^f  {\b P}_{A-1} \alpha_{A-1},\b p_1  \alpha_N  | {\hat J_\nu^A(0)} |
\alpha_A\b P_A\rangle ~,
\label{hadrontz}
\end{eqnarray}
where  $\alpha_i$ denotes the set of discrete quantum numbers of systems
$A$, $A-1$ and $N$. In Eq. (\ref{hadrontz})
 the vector $|\alpha_N {{\b p}_1}, \alpha _{A-1} {\b P}_{A-1}  E_{A-1}^f \rangle$
consists asymptotically of a nucleus $A-1$,
with momentum ${\b P}_{A-1}$ and intrinsic excitation
energy $E_{A-1}^f$,
and a nucleon with momentum ${{\b p}_1}$.
Two relevant experimentally measurable quantities which characterize
the process are
   the
{\it missing momentum}\,\, ${\b p}_m$  (i.e.   the  momentum of the $A-1$ system),
and  the\,\, {\it missing energy} \,\, $E_m$ defined, respectively, by
\be
{\b p}_m = {\b q} - {{\b p}_1} \,\,\,\,\,\,  E_m=
\sqrt{P_{A-1}^2}+M_N -M_A =  M_N + M_{A-1} -M_{A} + E_{A-1}^f =E_{min} +  E_{A-1}^f,
\label{missing}
\ee
where $E_{min}= E_{A}- E_{A-1} =  M_N + M_{A-1} -M_{A}$, and
 the (positive) ground-state energies
of $A$ and $A-1$ are denoted by $E_A$ and  $E_{A-1}$, respectively.
The exclusive cross section
can then be written  in  the well-known form
\be
{d^6 \sigma \over  d \Omega ' d {E'} ~ d^3{\b p}_m} =
\sigma_{Mott} ~ \sum_i ~ V_i ~ W_{i}^A( \nu , Q^2, {\b p}_m, E_m),
\label{2}
 \ee
where
$i \equiv\{L, T, LT, TT\}$, and $V_L$, $V_T$, $V_{LT}$, and $V_{TT}$ are well-known
kinematical factors.

\begin{figure}[!htp]
\centerline{
      \epsfysize=4cm\epsfbox{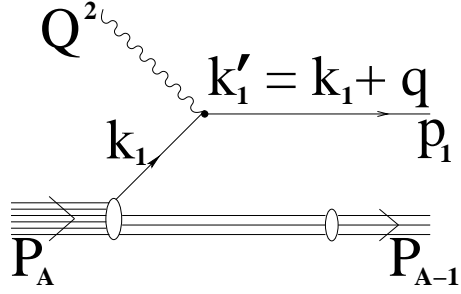}}
      \vspace{.5cm}
\centerline{\large\textbf{a)}\hspace{1cm}\normalsize}
      \vspace{.5cm}
\centerline{
      \epsfysize=4cm\epsfbox{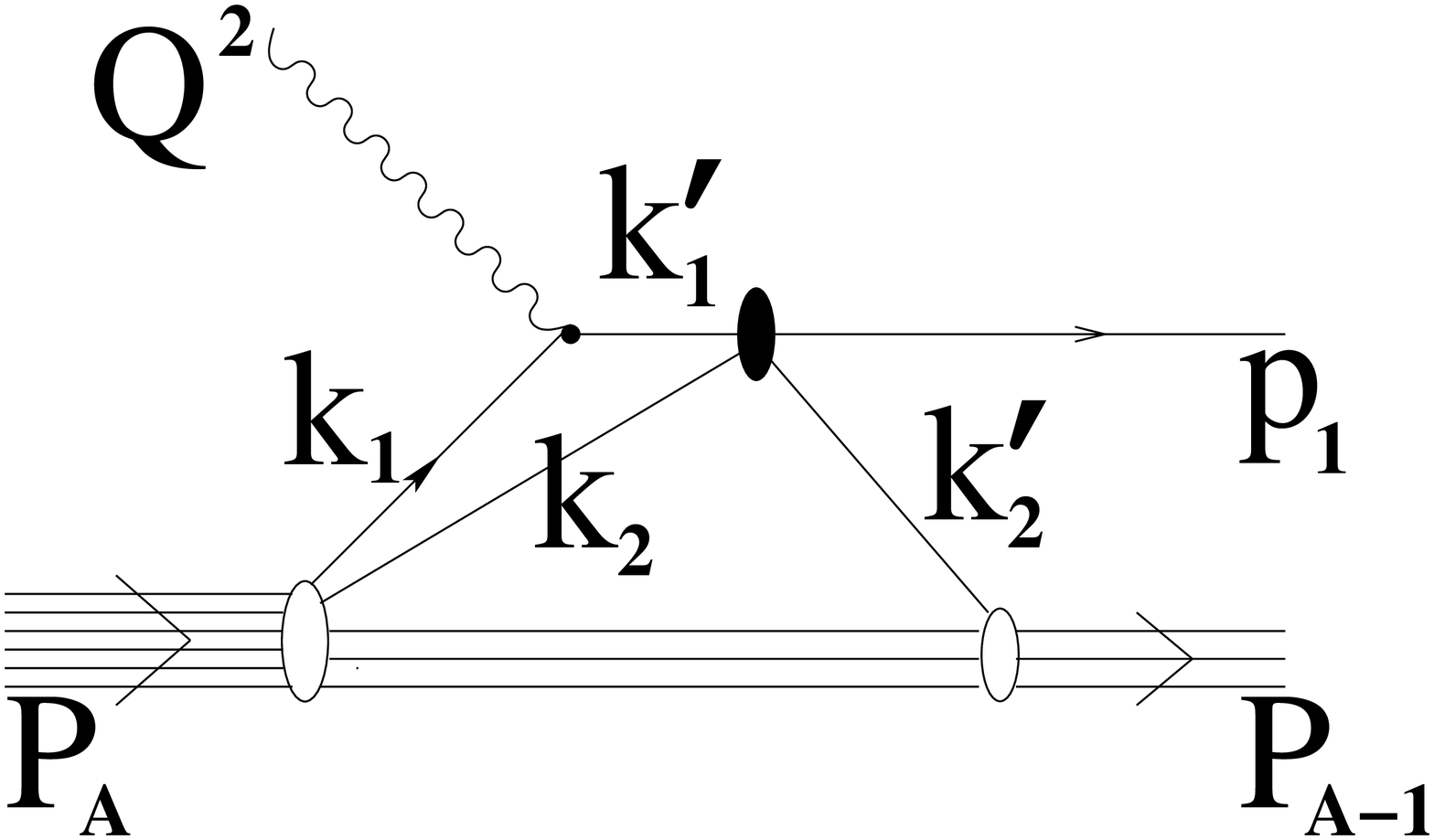}}
      \vspace{.5cm}
\centerline{\large\textbf{b)}\hspace{1cm}\normalsize}
      \vspace{.5cm}
\centerline{
      \epsfysize=4cm\epsfbox{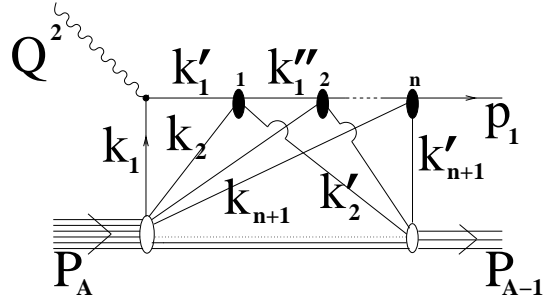}}
      \vspace{.5cm}
\centerline{\large\textbf{c)}\hspace{1cm}\normalsize}
\caption{ Feynman diagrams for the process $A(e,e^\prime p)(A-1)$. (a) describes
 the Plane Wave Impulse Approximation
(PWIA); (b)  the single rescattering; (c) the full
 $A-1$ rescattering. The
four-momenta of particle
$i$ before and after rescattering are  denoted by  $k_i$, $k_i^{'}$,  $k_i^{''}$,
etc.,
respectively. The  black oval spots denote the elastic
nucleon-nucleon  scattering
matrix..}
      \label{fig2}
\end{figure}

The evaluation of the nuclear response functions $W_{i}^A$ requires
the knowledge of  the nuclear vectors
$|\alpha_A\b P_A\rangle$ and  $|\alpha_N{{\b p}_1} , \alpha_{A-1}{\b P}_{A-1}  E_{A-1}^f \rangle$
 and the nuclear current
operators ${\hat J_\mu^A}(0)$.
Nowadays, there is no rigorous
quantum field theory to describe, from first principles,
a many body hadronic system, and one is forced to adhere to various approximations.
Whereas at  relatively low energies a consistent non relativistic treatment
of the electro-disintegration of   two- and three-body systems
can be pursued,  with increasing energy the treatment of the  three- body final state
 requires proper approximations. In the present  paper we describe  the two- and
three-body ground states in terms of realistic wave functions generated by
modern two-body interactions  ~\cite{pisa},
and treat the final state interaction by a diagrammatic approach of  the
elastic rescattering of the
struck nucleon with the nucleons of the $A-1$ system.  The
relevant diagrams  which, within such an approximation,
replace the One-Photon-Exchange
diagram of Fig. 1, are shown in Fig.~\ref{fig2}: the first
one represents the {\it Plane Wave Impulse Approximation} (PWIA), whereas the other ones
the final state rescattering (FSI).

 Although the PWIA appears to have a limited range of validity, it is
 useful to analyze its predictions
 since, within such an approximation, the cross
 section  is directly related  to a quantity, the $\it Spectral \,Function$,
  which, in the  case of few-body systems,
  can be calculated
 with high degrees of accuracy (see  \cite{sauer,cps,kps,cpslec}). The relevant point  here
 is that, provided   the FSI
  of the struck nucleon with the $A-1$ system can be disregarded,
   the Spectral Function yields direct information on
  the nuclear wave function.
 For such a reason,  we will present our results obtained
  within two distinct approaches:

  1. the {\it PWIA} (Fig.~\ref{fig2}a)), when  the struck proton is
  described by a plane wave, whereas the $A-1$ system in the final state,
  with momentum $P_{A-1}$,  represents the bound or continuum
  state solutions  of the Schr\"odinger equation with the same potential used to obtain
  the A-body wave function (note that some authors call PWIA the state in which
 {\it  all} particles in the continuum are described by plane waves);

2. the {\it full FSI approach} (Fig. \ref{fig2}b,c)), when the
$A-1$ system (in the ground or continuum states) is still described by  the exact
solution of the Schr\"odinger equation, and  the  interaction of the
 struck nucleon with the $A-1$ nucleons
 is treated  by evaluating the Feynman diagrams of Fig. 2,  either in the
GA or the GEA approximations.

\section{The Plane Wave Impulse Approximation and the Nuclear Spectral Function}
\label{susecA}

The main merit of the PWIA is that it allows one  to express the $\it nuclear$ response functions $W_{i}^A$  in terms  of
the $\it nucleon$ response functions which are very well known from $e-N$ experiments.
As a matter of fact, by expressing the hadronic tensor for the nucleus $A$
(Eq. (\ref{hadrontz}))
in terms of the  hadronic tensor   for the nucleon $N$
 \be
 W_{\mu\nu}^N&=&  { 1 \over 4 \pi M_N}\overline{\sum_{\alpha_N}}  \sum_{\alpha_N'}
(2\pi)^4 \delta^{(4)} (p + q - p_1)   
\langle \alpha_N {\b p}_1| {\hat J_\mu^N}(0) |
\alpha_N^{'}{\b p}_1^{'}  \rangle \langle  \alpha_N^{'}{\b p}_1^{'}|{\hat J_\nu^N} (0) |
\alpha_N{\b p}_1\rangle,
\label{nuctz}
\ee
the cross section assumes the following form  (see e.g. Refs ~\cite{PWIA1,PWIA2,forest})
\be
&&
\frac{d^6\sigma}{dE'd\Omega 'd{\bf p}_m}=
      K({ Q}^2,x,{\bf p}_m)\, \sigma^{eN}({\bar Q}^2,{\bf p}_m)
P_A({|\b k}_1|, E),
\label{eq4}
\ee
where ${\bar Q}^2$ =${\b q}^2 - {\bar q}_0^2$
(\, $\bar{q}_0 = q_0 + M_A -
\sqrt {({\b k}_1^2 + (M_{A-1}^f)^2} -
\sqrt{ {{\b k}_1}^2 + M_N^2}\,\,$), and
$K({ Q}^2,x,{\bf p})$   a kinematical factor. In Eq. (\ref{eq4}),
$\sigma^{eN}({\bar Q}^2,{\bf p}_m)$  is the  cross section
 describing electron scattering by an off-shell nucleon,
$x={Q}^2/2M_N q_0$  is  the Bjorken scaling variable,
 ${\b k}_1=-{\bf p}_m$ is the
  nucleon momentum before interaction,
 $E \equiv E_{mis} = E_{min} + E_{A-1}^f$ is
 the
\textit{removal energy} and
 $P(|{\bf k}_1|,E)$ is the nucleon Spectral Function, which can be written as follows:
\be
P({|{\b k}_1|},E)&=& \frac {1}{(2\pi)^3} \frac{1}{2J_A+1}\,\,\, \sum_{f}\,\,\,
\sum_{{\cal M}_A,\,{\cal M}_{A-1},\,\sigma_N}
\left | \langle \alpha_A \b P_A|
\alpha_N{{\b k}_1} , \alpha_{A-1}{\b P}_{A-1}  E_{A-1}^f \rangle  \right |^2 \times\\\nonumber
&\times&\delta \left (E - (E_{A-1}^f + E_{min}) \right),
\label{eq5}
\ee

\noindent where  ${\cal M}_A$,  ${\cal M}_{A-1}$, and $\sigma_n$, are the spin projections,
and  the sum over $f$  includes all possible discrete and continuum states of the
 $A-1$ system.

Whereas   the Spectral Function for the Deuteron (D) has a 
  particularly simple form, viz.

 \be
 P_D(|{\bf k}_1|, E) = n_D(|{\bf k}_1|) \delta (E-\epsilon_D),
 \label{deut}
 \ee
 where $\epsilon_D$ is the (positive) binding energy of the deuteron and
 $n_D (|\b{{\b{k}}_1}|) =
  {(2\pi^2)}^{-1}\left( u_S^2( |\b{{\b{k}}_1}|) + u_D^2( |\b{{\b{k}}_1}|)\right)$
 the nucleon momentum distribution, in the case of A=3,
the proton Spectral Function consists of two parts,
\be
 P_{He}(|{\b k}_1|,E) =  P_{gr}(|{\b k}_1|,E) + P_{ex}(|{\b k}_1|,E),
\label{eq13}
\ee

\noindent
The first one, or {\it ground} part $P_{gr}$,  has the following form
\be
P_{gr}(|{\b k}_1|,E) = n_{gr}(|{\b k}_1|)\delta(E - E_{min}),
\label{pgr}
\ee
where $E_{min}= |E_3|-|E_2| \approx 5.49 MeV$, and
$ n_{gr}(|{\b k}_1|)$, which corresponds to the two-body break-up (2bbu) channel
 $^3He\to D+p$,  is  (hereafter, the projection of the spin of nucleon $i$ will be denoted
 by $s_i$)
\be
n_{gr}(|{\b k}_1|)=\frac{1}{(2 \pi)^3} \frac{1}{2}
  \sum_{\CM_3, \CM_2,s_1}
  \left | \int {\rm e}^{-i\boldrho{\b k}_1} \chi_{\frac12 s_1}^\dagger
  \Psi_{D}^{{\CM_2}\dagger}(\b{r} )
  \Psi_{He}^{\CM_3}(\boldrho,\b{r})
  d \boldrho d {\bf r}\right |^2.
\label{ngr}
\ee

\noindent In Eq. (\ref{ngr}) $\Psi_{He}^{\CM_3}(\boldrho,\b{r})$ is the  $^3He$
  wave function,
${\cal M}_3$  the projection of the  spin of  $^3He$ , and  $\b{r}$ and
$\boldrho$ the Jacobi coordinates describing, respectively, the motion of
the spectator
pair and the motion of the struck (active) nucleon with respect to the CM of the pair.

  The second, or {\it excited}  part $P_{ex}$, of $P_{He}(|{\b k}_1|,E)$,
  corresponds
to the three-body break-up (3bbu) channel $^3He\to (np)+p$
and  can be written as follows
 \be
 P_{ex}(|{\b k}_1|,E)&=& \frac{1}{(2 \pi)^3} \frac{1}{2}
  \sum_{{\cal M}_3, S_{23},s_1}
       \int \frac{d^3 \b {t}}{(2\pi)^3}
  \left | \int  {\rm e}^{-i\boldrho{\b k}_1} \chi_{\frac12 s_1}^\dagger
 \Psi_{np}^{\b{t}\dagger}(\b{r}) \Psi_{He}^{{\cal M}_3}(\boldrho,\b{r})
    d\boldrho d {\bf r}\right |^2\times\nonumber\\
&\times&\delta \left( E - \frac{\b {t}^2}{M_N} - E_3 \right),
\label{piex}
\ee
where   $\Psi_{np}^{\b{t}}(\b{r})$ is  the two-body continuum wave function
 characterized by spin projection $S_{23}$ and by the relative momentum
${\b t}= \frac {{\b k}_2 - {\b k}_3}{2}$
of the $np$ pair in the continuum.
 Obviously, for the neutron Spectral Function, only the excited part  (\ref{piex})
 contributes.

\begin{figure}[!htp]                     
\epsfig{file=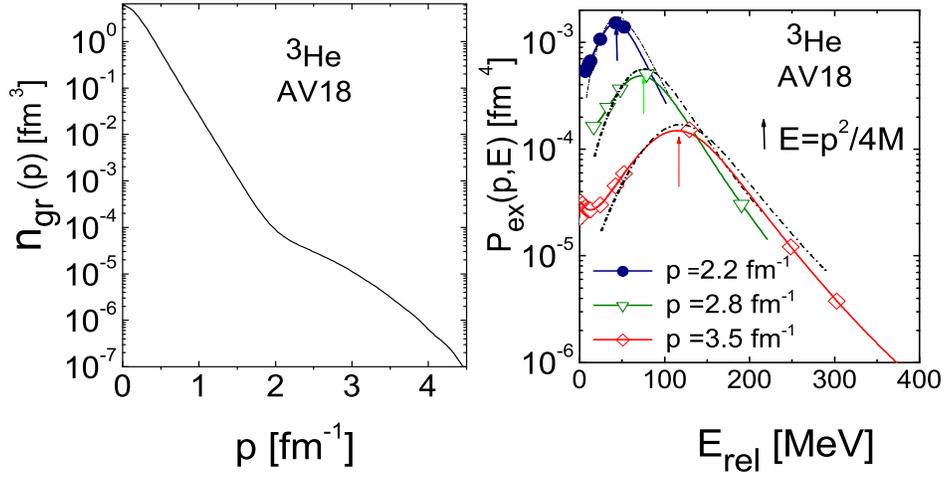,width=14.0cm,height=8.0cm}
\caption{The  proton Spectral Function of $^3He$ (Eq.(\protect\ref{eq13})).
 {\it Left panel}: $n_{gr}$ (Eq. (\ref{ngr})) {\it vs}  $p\equiv|{\b k}_1|$.
 {\it Right panel}: $P_{ex}$ (Eq. (\ref{piex})) {\it vs}  the excitation energy of
 the two-nucleon system in the continuum
  $E_{rel}=\displaystyle\frac{\b {t}^2}{M_N}$ = $E_2^f = E- E_{min}$,  for various
  values of  $p\equiv|{\b k}_1|$.  The dot-dashed curves represent the Plane Wave Approximation
  (PWA), when the three particles in the continuum are described by plane waves,
  whereas  the full curves correspond to the PWIA, when the interaction in the
  spectator proton-neutron pair is taken into account.  The arrows indicate
 the position of the peak ($\sim p^2/4M_N$) predicted by the two-nucleon
  correlation
 model for the Spectral Function \cite{fscs} (three-body wave function  from
  Ref. \cite{pisa}, $AV18$ interaction \cite{av18}).}
      \label{fig3}
\end{figure}

\begin{figure}[!htp]
   \includegraphics[width=8cm,height=9.5cm]{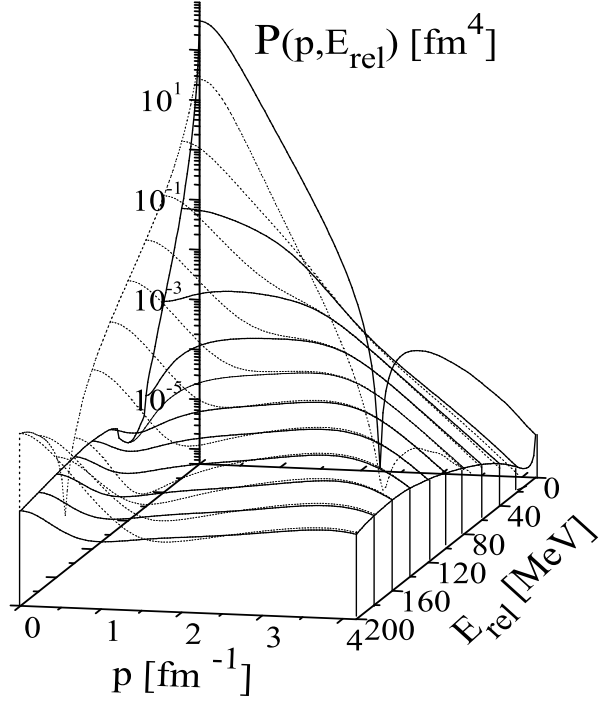}
\caption{ The neutron Spectral Function
of $^3He$ (Eq. (\ref{piex})) {\it vs} $p\equiv|{\b k}_1|$ and
 the excitation energy of
 the two-nucleon system in the continuum
  $E_{rel}=\displaystyle\frac{\b {t}^2}{M_N}$ = $E_2^f = E- E_{min}$.
   The dotted curves represent the
  $PWA$, when the three particles in the continuum are described by plane waves,
  whereas  the full curves correspond to the $PWIA$, when the interaction in the
  spectator proton-proton pair is taken into account  (three-body wave function  from
  Ref. \cite{pisa}, $AV18$ interaction \cite{av18}).}
\label{fig4}
\end{figure}

   In Fig.~\ref{fig3} we  show the Spectral Function of  $^3He$  obtained
   using   the
  variational three-body wave function  by the Pisa group ~\cite{pisa}
  corresponding to  the
   realistic AV18 potential~\cite{av18} (see   Appendix A).
  The two-body wave function entering Eq. (\ref{piex}) has
   been obtained by solving the Schr\"odinger equation for the
  continuum using the same $AV18$ two nucleon potential.
  Our results for the Spectral Function agree with the ones obtained in
  Ref. \cite{kps},  where the same three-body wave function has been used.
  The normalization of the Proton Spectral Function has been fixed to 2
  (two protons) and  the normalization of  the neutron Spectral Function  to one.
 In Fig.~\ref{fig3} we  also show  the results predicted by the
Plane Wave Approximation (PWA), which corresponds to the replacement
 of the continuum interacting $(n-p)$ pair wave function with two plane waves.
    The left panel of  Fig.~\ref{fig3}
    shows $n_{gr}$ and the right panel  $P_{ex}$.
   It can be seen that: i) $P_{ex}$
  exhibits  maxima centered
  approximately  at $E_m\sim\displaystyle\frac{{{\b k}_1}^2}{4M_N}$, ii)
  around these values of $E_m$ and ${\b k}_1$ the
Spectral Functions, calculated disregarding the interaction
in the $NN$-pair in the continuum  (PWA)  and
  taking it into account (PWIA), are almost identical,
   in agreement with the results obtained
  long ago \cite{cpslec} with the Spectral Function corresponding to
  the Reid Soft Core Interaction \cite{rsc}. The region centered at
   $E_m\sim\displaystyle\frac{{{\b k}_1}^2}{4M_N}$
  is the so-called two-nucleon correlation region~ \cite{FSreport}, when
  one of the nucleons of the spectator $NN$-pair is fast, the other one
  being basically at rest (
  for an improved description which takes
  into account the motion of the third, uncorrelated nucleon, or the
   $A-2$ spectator system in case of heavier nuclei, see \cite{fscs}). 
  Then the fast nucleon becomes strongly correlated
  with the active nucleon (the proton,
   in the case of the proton Spectral Function,
  or  the neutron, in  the case of the neutron Spectral Function)  forming
   a correlated
  pair which carries  most of the nuclear momentum.  In this case, it is
  intuitively expected that the slow nucleon acts
  as a passive spectator and, consequently, only the interaction
  in the correlated pair can be  relevant for the Spectral Function.
  Hence, in this region the calculations including or omitting
   the interaction in the spectator pair, are expected to  provide essentially
  the same results, as confirmed by  present
   and previous  calculations of the Spectral Function
   \cite{greno,pavia,cpslec}. The situation which has been just described, is
   clearly  illustrated in   Fig.~\ref{fig4}, where the three-dimensional
  neutron Spectral Function is presented.

  The PWIA results suggest that experimental insight about the structure of the
  nuclear wave function at short distances can be  obtained from
 $A(e,e'p)(A-1)$ processes, provided the  PWIA  entirely exhausts
  the reaction mechanism. Unfortunately, we know that
  in many  cases the simplified
  PWIA mechanism fails to describe the experimental data (see, e.g. a
  recent discussion in Ref.~\cite{gleockle_SF}). However, a properly
  chosen kinematics can still leave room for studying  $NN$
  correlations.  It is clear, from Figs.~\ref{fig3}~and~\ref{fig4},
  that such a kinematics should  be located
  around the two-nucleon correlation region, in order to exclude
  the influence of the final state interaction between the spectator nucleons.
  This requires  high values of  the missing energy and  momentum of the active
  nucleon  (see Fig.~\ref{fig4}).
  Another important condition is that the range of $|\b{q}|$ and  $ q_0$
  should   not be too far from the quasi-elastic peak, where
  $x \simeq 1$.  In this case the corrections  from the off mass shell effects and meson
  production are minimized, and only the final state interaction of the
  hit nucleon with the spectators   becomes relevant.
  However, it should  always be  kept in mind, that if a region exists
  where the interaction {\it in} the spectator
   pair (the $A-1$ system in case of complex nuclei) can be neglected,
    this is no guarantee  that the interaction of the struck nucleon {\it with} the
  nucleons of the spectator pair (the $A-1$ system), can be neglected as well.
  It is clear therefore that one has to
 go beyond the PWIA, which is precisely the aim of the present paper.
 The effects of the full FSI  on the process $^3He(e,e'2p)n$ have
 been recently investigated,  treating
 the FSI within the GA \cite{claleo2p}. In the present paper we will investigate
 the same topic in $A(e,e'p)X$ process off $^2H$ and $^3He$ within both the GA and
the GEA.

    \section{The full Final State Interaction within the Generalized Eikonal Approximation
    }
 \label{FSI}

Let us consider the interaction of the incoming virtual photon, $\gamma^*$,
with a bound nucleon (the active nucleon)
of low virtuality ($p^2\sim M_N^2$) at a kinematics not very different
from the quasi-elastic one, i.e.  corresponding to $x\sim 1$. In the
 quasi-elastic kinematics,
 the virtuality  of the struck nucleon after
    $\gamma^*$-absorption  is also rather low and,
    provided ${\b p}_1$  is sufficiently high, nucleon rescattering
    with the "spectator" $A-1$
    can be   described to a large extent in terms of  multiple
 elastic scattering processes in the forward direction
 (in the system of reference  where the target nucleon is at rest).
  These rescattering processes are diagrammatically
   depicted in Fig.~\ref{fig2} where, as in the rest of this paper,
    the internal and intermediate state
   momenta are  denoted by $k_i$'s and the final state momenta by $p_i$'s.
    The diagrams essentially describe the process of multiple scattering
   in the most general case, within  the assumption that all intermediate
    nucleons are on-shell.
   The low virtuality (before and after $\gamma^*$
   absorption)  of the active nucleon, coupled with  the forward
   propagation,  allows one to simplify the description of the final
    state interaction, which can be
   treated within  the
  eikonal approximation.
  Before illustrating  in detail the approach we have used
  to treat  FSI in $(e,e'p)$ reactions,
  we would like to discuss an important related issue (see also \cite{factor}), namely
  the validity of the factorization approximation, frequently
  used in calculations at
  high $Q^2$,  and consisting in factorizing the
  $(e,e'p)$
  cross section into an e.m. and a nuclear parts, in spite of the fact
  that factorization, holding exactly in PWIA, is violated when FSI is
  taken into account. In the next Section the factorization approximation will be
   discussed
  within  the GA and the  GEA.

\subsection{The FSI in $A(e,e'p)B$ processes within a diagrammatic approach}

\begin{figure}[!htp]
\centerline{
      \epsfysize=5cm\epsfbox{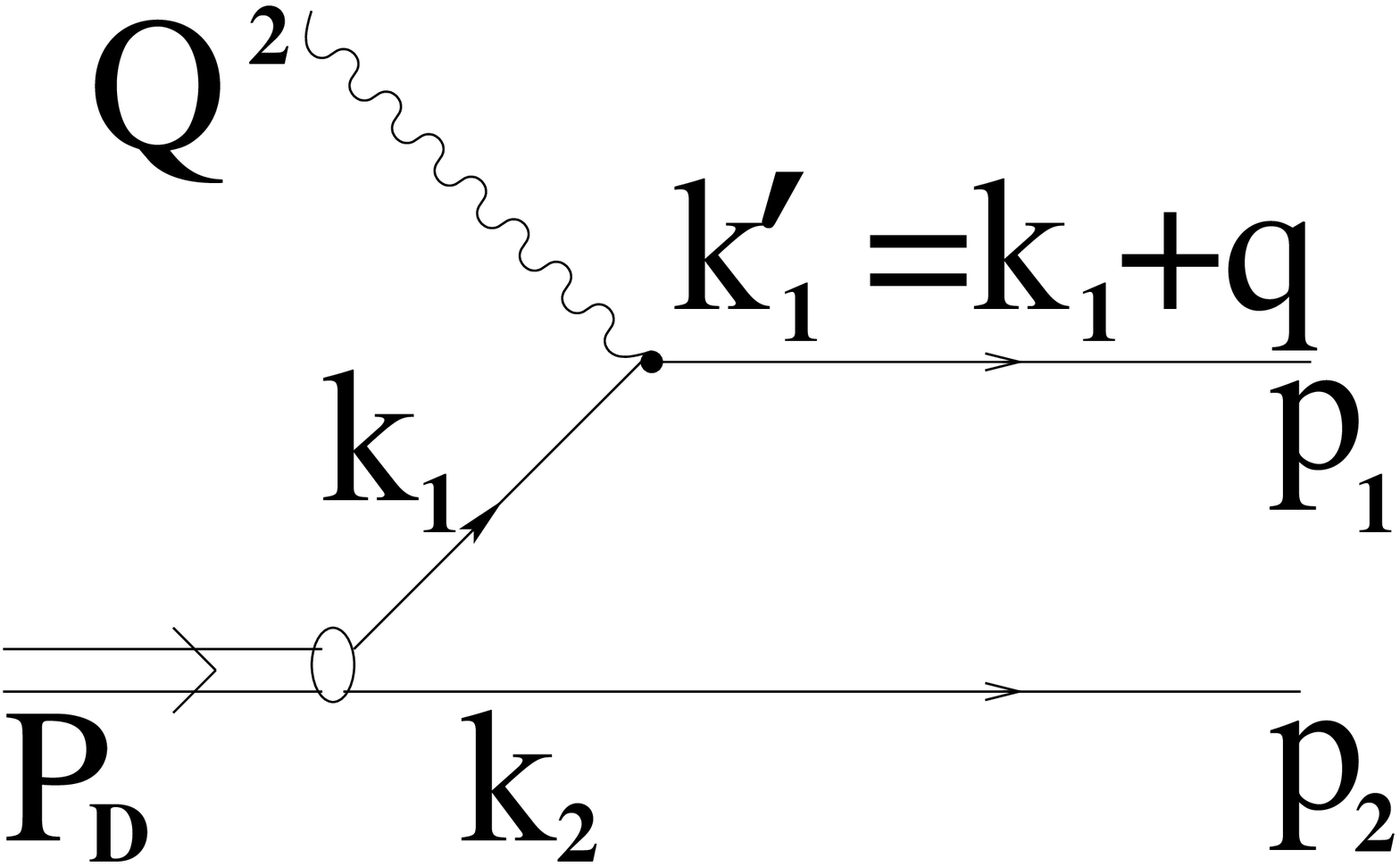}}
      \vspace{.5cm}
\centerline{\large\textbf{a)}\hspace{1cm}\normalsize}
      \vspace{.5cm}
\centerline{
      \epsfysize=5cm\epsfbox{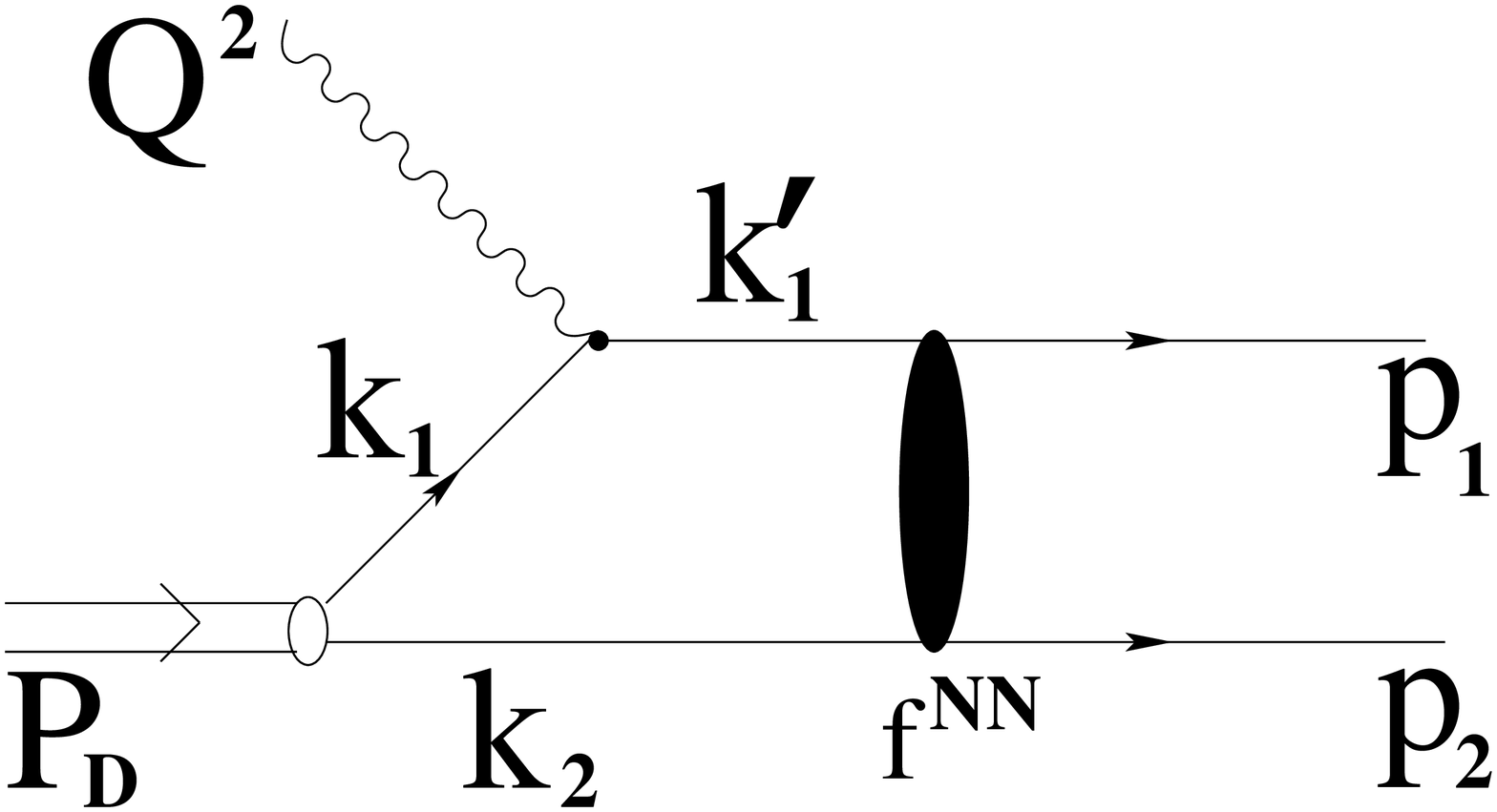}}
      \vspace{.5cm}
\centerline{\large\textbf{b)}\hspace{1cm}\normalsize}
\caption{The Feynman diagrams for the process $^2H(e,e^\prime p)n$ representing
 the PWIA) $(a))$, and the single $(b))$
      rescattering in the final state. $f^{NN}$ denotes the elastic
      NN scattering amplitude.}
      \label{fig5}
\end{figure}

  Most of the problems one faces   when trying
  to develop  a fully covariant  treatment of FSI,  arise because of
  the hadrons' spins. Therefore,
  let us rewrite  the hadronic tensor (Eq. (\ref{hadrontz}))
 in the following, fully equivalent form, which however exhibits explicitly
  the dependence upon the spin quantum numbers
 \begin{eqnarray}
W_{\mu\nu}^{A} &=& { 1 \over 4 \pi M_A} {\overline {\sum_{\alpha_A}}} \sum_{ \alpha_{A-1},s_1}
 T_\mu^{\dagger}({\cal M}_A,{\cal M}_{A-1},s_1)T_\nu({\cal M}_A,{\cal M}_{A-1}, s_1) 
  (2\pi)^4\delta^{(4)} (P_A + q - P_{A-1} -p_1),\nonumber\\&&
\label{hadrontz1}
\end{eqnarray}
where $T_\mu$ is a short-hand notation for the transition matrix element
\be
 T_\mu({\cal M}_A,{\cal M}_{A-1},s_1)
 \equiv \langle \alpha_{A-1}{\b P}_{A-1}  E_{A-1}^f ,s_1 \b p_1  | {\hat J_\mu^A}(0) |
\alpha_A\b P_A\rangle.
\label{trans}
\ee
The basic assumption underlying the eikonal diagrammatic
treatment of FSI  at high $Q^2$ is that  the transition matrix
element $T_\mu$ for  a nucleus $A$
 can be written in the following form
\be
T_\mu({\cal M}_A,{\cal M}_{A-1},s_1) =
\sum_{n=0}^{A-1} T_\mu^{(n)}({\cal M}_A,{\cal M}_{A-1},s_1),
\label{ti}
\ee
where  the superscript $(n)$ corresponds to the order of rescattering
of the struck particle with the $A-1$ nucleons (the "spectator" nucleons),
 namely $T_\mu^{(0)}$ corresponds to the PWIA (no rescattering),
 $T_\mu^{(1)}$ to the single rescattering of the struck nucleons with the spectator ones,  $T_\mu^{(2)}$
 to double rescattering, and so on.
Such an approach is expected to be valid either  at high energies,  when
particles propagate mostly in the forward direction along the direction of
the three-momentum transfer ${\bf q}$, or when the momentum of the struck nucleon ${\bf p}_1$
relative to $A-1$ is sufficiently high; in both cases
 the eikonal approximation could be applied.
 The calculation of the rescattering part of $T_\mu$ in terms of
 Feynman diagrams
 appears in principle to be  a prohibitive
  relativistic task due, as previously stressed,
  to the treatment
  of the spin. A relevant simplification occurs if the cross
   section factorizes into
  the e.m. and the nuclear parts  and,  as a matter of fact, many calculations performed
   within the eikonal approximation
  treatment of the FSI, simply assume factorization. Let us try
  to analyze  the limits  of validity of such an assumption, and to this end
let us  consider  the deuteron.
   In this case the Feynman diagrams describing rescattering are given
in Fig.~\ref{fig5},  and the corresponding matrix element  is
 \be
T_\mu ({\cal M}_2,s_1,s_2)= T_\mu^{(0)} ({\cal M}_2,s_1,s_2)+T_\mu^{(1)} ({\cal M}_2,s_1,s_2),
\label{mult}
\ee
\noindent where
 ${\cal M}_2$, $s_1$ and $s_2$ are the spin projection
 of the deuteron,  and of nucleon
"1" (the active nucleon)
 and nucleon "2" (the spectator nucleon) in
the final state.  Eq. (\ref{mult}) obviously states
 that in the deuteron the interaction between the struck and the spectator nucleon
  can occur only via single rescattering.
 The cross section of the process is given by
\be
\frac{d^5\sigma}{dE' d\Omega'}=
\sigma_{Mott}\tilde l^{\mu\nu}L_{\mu\nu}^D\frac{d^3p_1}{(2\pi)^3 2E_1}
\frac{d^3p_2}{(2\pi)^3 2E_2},
\label{cross1}
\ee
where $E_i=\sqrt {{\b p}_i^2+M^2}$, and the hadronic tensor
is as follows
\be
L_{\mu\nu}^D&=&\frac{1}{2M_D}\frac13\sum\limits_{{\cal M}_2,s_1,s_2}
T_\mu^{\dagger}({\cal M}_2,s_1,s_2) T_\nu ({\cal M}_2,s_1,s_2)
(2\pi)^4\delta^{(4)}\left( P_D+q-p_1-p_2\right).
\label{tensor1}
\ee

  Let us now  obtain the factorization of the cross section, expressed in terms of
  the hadronic tensor (\ref{tensor1}),  within a fully covariant
  approach.

\subsubsection{The PWIA and the factorization of the cross section.}
 The PWIA for the process $^2H(e,e'p)n$ within a covariant Feynman
 diagram approach  has been considered  by various authors (see e.g.
 \cite{deuteronrel,gross,bs}).
 The matrix element  $T_{\mu}=T_{\mu}^{(0)}$, in such a case,
  has the following form
\be
&&T_\mu^{(0)} ({\cal M}_2,s_1,s_2)= \nonumber\\
&=&\dfrac{1}{2M_N}
\sum\limits_{\tilde{{s}_1}} J_\mu^{eN}(Q^2,p_1,k_1,\tilde{s_1},s_1)
\left [ \bar u (\b{k}_1,\tilde{s_1}) \Phi_{D}^{{\cal M}_2}(k_1,k_2)
 (\hat k_2+M_N) v(\b{k}_2,s_2)\right ],
\label{ampli}
\ee
where
\be
J_\mu^{eN}(Q^2,p_1,k_1,\tilde{s_1},s_1)
=\langle \b{p}_1,s_1| \Gamma_{\mu}^{\gamma^{*}N}
(Q^2,k_1^2) | \b{k}_1,\tilde{s_1}\ra,
\label{corrente}
\ee
 $\Gamma_{\mu}^{\gamma^{*}N}
(Q^2,k_1^2)$ is the e.m. vertex, and $\Phi_{D}^{{{\cal M}_2}}(k_1,k_2)$
the covariant deuteron amplitude corresponding to  the
$D \rightarrow (pn)$ vertex. The explicit form of the amplitude
 $\Phi_{D}^{{{\cal M}_2}}(k_1,k_2)$ depends upon the specific covariant
 model used to describe the deuteron and could be found
 elsewhere (see e.g.,~\cite{deuteronrel,gross,amplBS}). Here, without loss of generality,
 we will use the Bethe-Salpeter (BS) formalism  according to Refs.
 \cite{amplBS} and \cite{bs}.

When Eq. (\ref{ampli}) is placed  in Eq. (\ref{tensor1}),
 the e.m. and nuclear parts  gets  coupled by the summation over the intermediate
spins $\tilde s_1$ and ${\tilde {s_{1}}}'$.
However, it can be shown (see Appendix~\ref{app:fact})
that  the square of the expression in brackets in Eq.(\ref{ampli}) after summation
over ${\cal M}_2$ and $s_2$  yields a $\delta$ function
$\delta_{ \tilde s_1 {\tilde s_{1}}^{'}}$, i.e. becomes diagonal in
 $\tilde s_1$; this leads to the decoupling between
 the e.m and the nuclear parts in Eq.(\ref{ampli}), with the resulting
  hadronic tensor given by
 \be
L_{\mu \nu}^D&=&
\frac{1}{2M_D}\frac13\sum_{{\cal M}_2, s_1,s_2}T_{\mu}^\dagger ({\cal M}_2,s_1,s_2)
T_{\nu}({\cal M}_2,s_1,s_2)=\nonumber\\
&=& 2M_D\left( 2E_{\b {p}_1} 2E_{\b {k}_1}L_{\mu\nu}^{eN}(Q^2,p_1,k_1)
 \right )\, n_D(|{\b k}_1|)\,
 (2\pi)^4\delta^{(4)}\left( P_D+q-p_1-p_2\right).
\label{eq31}
\ee

\noindent In Eq. (\ref{eq31}),  $n_D$ is  the deuteron momentum distribution
given by 
\be
n_D(|\b{k}_1|) &=&\dfrac13\sum\limits_{\CM_2,\tilde s_1,s_2}
\left |\left [ \bar u (\b{k}_1,\tilde{s_1}) \Phi_{D}^{{\cal M}_2}(k_1,k_2)
 (\hat k_2+M_N) v(\b{k}_2,s_2)\right ] \right |^2 =\nonumber\\
 &=&
\dfrac13\sum\limits_{\CM_2,\tilde s_1,s_2} \left | \left\langle
 \tilde s_1, s_2 | \Psi_D^{\CM_2}(\b{k_1})
\right\rangle \right |^2
= \frac{1}{2\pi^2}\left (u_S^2( |\b{{\b{k}}_1}|) + u_D^2( |\b{{\b{k}}_1}|)\right ),
\label{momdis}
\ee
\noindent where   the
 (covariant) deuteron wave function has been  cast in a form similar to the
  non relativistic one
 with
the scalar parts of the wave function, $u_L(|{\bf k}|)$'s,  related to
 the corresponding vertex functions, $G_L(k_1^2,k_2^2=M_N^2)$,
  by a well known definition (see Eqs. (B8) and (B9)) leading to
 \be
 u_L(|{\bf k}|) \sim
\sqrt{2M_N} \frac{G_L\left (|{\bf k}|,k_{10}=M_D-E_{\bf {k}}\right)}{k^2-M_N^2}.
\label{vertex}
\ee
Placing Eq. (\ref{eq31}) in Eq. (\ref{cross1}) the well known factorized form
for the cross section is obtained, {\it viz.}
\be
\frac{d^6\sigma}{d E' d\Omega' d{\b p}_m}=
K({ Q}^2,x,{\bf p}_m)\, \sigma^{eN}({\bar Q}^2,{\bf p}_m)
 {n}_D(|{\b{k}}_1|))
 \delta (q_0 + M_D - E_{{\b{k}}_1 + {\b{q}}}-E_{{\b{k}}_1}).
 \label{eq19}
\ee

We reiterate that   factorization  has been obtained because
  the sum over $s_2$
and ${\cal M}_2$  in (\ref{tensor1})  leads to the appearance of a delta
 function $\delta _{{\tilde s_1}',\tilde s_1}$,
 which means, in turn, that  the square of $T_\mu^{(0)}$ becomes diagonal in
${\tilde s_1}$. This  particular (exact) result is  part of a more general assertion
 that within the PWIA the nuclear Spectral Function
is always diagonal in spins \cite{forest}. Let us now consider FSI; in this case
 the tensor (\ref{tensor1})
is off-diagonal in spins and factorization does not occur.
However, we will show  that under certain kinematical
conditions, satisfied to a large extent by the  GA and GEA, factorization can be recovered.

 \subsubsection{FSI: the single scattering contribution
  and factorization of the cross section.}

Let us compute the second diagram of Fig. \ref{fig5}. To this end,
we
introduce a two-nucleon scattering  operator $\hat T$ in terms of  which
  the elastic scattering amplitude $f^{NN}$,
describing the elastic scattering of two on-shell nucleons,  will  be defined as follows
 \be
 f^{NN}_{\tilde{s_1}\tilde s_2;s_1,s_2}(\b{p}_1,\b{p}_2;\b{k}_1,\b{k}_2)
=\bar u(\b{p}_1,s_1) \bar u(\b{p}_2,s_2)\ \hat T\
 u(\b{k}_1,\tilde{s_1})  u(\b{k}_2,\tilde{s_2}),
 \label{ampiezza}
 \ee

 \noindent which is obviously the free $NN$ scattering amplitude; for a bound nucleon one
  has in principle
 to consider off-shell effects but  in the GEA no virtuality is considered; this
  could be done for example by the approach
   of  Ref. \cite{misha},
  by introducing cut-off form factors in the
 corresponding nucleon lines,
 which formally leads to two Feynman diagrams with different "nucleonic" masses.
 In presence of FSI, the transition matrix element  is
  \be
T_\mu ({\cal M}_2,s_1,s_2)= T_\mu^{(0)} ({\cal M}_2,s_1,s_2)+T_\mu^{(1)} ({\cal M}_2,s_1,s_2)
\label{mult1}
\ee

\noindent with    $T_\mu^{(0)}$ given again by Eq. (\ref{ampli}), and
  $T_\mu^{(1)}$ given by
 the following  form (note that  henceforth  we always have
 ${\bf k}_1 = -{\bf k}_2$)
 \be &&
 T_{\mu}^{(1)} ({\cal M}_2,s_1,s_2)=
\dfrac{1}{2M_N} \sum\limits_{\tilde{s_1}\tilde{s_1}^\prime\tilde {s_2}}
\int \dfrac{d^4 k_2}{i(2\pi)^4}
\frac{ f^{NN}_{\tilde{s_1}^\prime\tilde s_2;s_1,s_2}(\b{p}_1,\b{p}_2,\b{k}_1^\prime,\b{k}_2)}
{k_1'{^2}-M_N^2+i\varepsilon}
 \times\nonumber\\
&\times&\left [ \bar u(\b{k}_1^\prime,\tilde{s_1}^\prime)
 \Gamma_{\mu}^{\gamma^{*}N}(Q^2,k_1^{\prime 2}) u(\b{k}_1,\tilde{s_1})\right]
  \left [ \bar u(\b{k}_1,\tilde{s_1})
 \Phi_{D}^{{\cal M}_2}(k_1,k_2) v(\b{k}_2,{\tilde s_2})\right].
 \label{eq38a}
 \ee

\noindent The full matrix element will therefore be
\be
      &&T_{\mu}({\cal M}_2,s_1,s_2) =\nonumber\\
      &=& \dfrac{1}{2M_N}
\sum\limits_{\tilde{{s}_1}} J_\mu^{eN}(Q^2,{\b p_1},{\b p}_m,\tilde{s_1},s_1)
\left [ \bar u (\b{k}_1,\tilde{s_1}) \Phi_{D}^{{\cal M}_2}(k_1,k_2)
 (\hat k_2+M_N) v(\b{k}_2,s_2)\right ]+\nonumber\\
 &+&
\dfrac{1}{2M_N} \sum\limits_{\tilde{s_1}\tilde{s_1}^\prime\tilde {s_2}}
\int \dfrac{d^4 k_2}{i(2\pi)^4}
\frac{ f^{NN}_{\tilde{s_1}^\prime\tilde s_2;s_1,s_2}(\b{p}_1,\b{p}_2,\b{k}_1^\prime,\b{k}_2)}
{k_1'{^2}-M_N^2+i\varepsilon}
 \times\nonumber\\
&\times&\left [ \bar u(\b{k}_1^\prime,\tilde{s_1}^\prime)
 \Gamma_{\mu}^{\gamma^{*}N}(Q^2,k_1^{\prime 2}) u(\b{k}_1,\tilde{s_1})\right]
  \left [ \bar u(\b{k}_1,\tilde{s_1})
 \Phi_{D}^{{\cal M}_2}(k_1,k_2) v(\b{k}_2,{{\tilde s}_2})\right].
 \label{ttot}
 \ee

When  Eq. (\ref{ttot}) is placed into Eq. (\ref{tensor1}),
  the resulting equation is not diagonal in the
spin quantum numbers and factorization
does not hold. Let us however consider the basic assumptions
underlying the eikonal multiple scattering approach, viz.:
\begin{enumerate}
\item the momentum transfer ${\boldkappa}$ in the elastic
rescattering is small and mostly  transverse i.e.
\be
{\boldkappa} = {\b p}_1 - {\b k}_1^{'} = {\b k}_2 - {\b p}_2 \simeq
{{\b k}_2}_\perp
- {{\b p}_2}_\perp=
{\boldkappa}_\perp
\ee
\item  the spin-flip part of the $NN$ amplitude is very
small, which means that, taking into account Point 1, one can write
\be
f^{NN}_{\tilde{s_1}^\prime\tilde s_2;s_1,s_2}(\b{p}_1,\b{p}_2,\b{k}_1^\prime,\b{k}_2)
 \approx \delta_{\tilde{s_1}^\prime, s_1}\delta_{\tilde s_2, s_2}
 f^{NN}(\boldkappa_\perp)
 \label{spinflip}
 \ee
 which  is realized  either at high  values of the
  three-momentum transfer $\b q$, or
at  high values of  the  momentum ${\b p}_1$ of the struck nucleon relative
to the $A-1$ spectator nucleons.

\end{enumerate}
If the above conditions are satisfied, Eq. (\ref{eq38a}) assumes the following
 form (cf. Appendix B)
 \be
  T_\mu^{(1)}({\cal M}_2,s_1,s_2)&\simeq&
\sum_{\tilde s_1}
J_\nu^{eN}(Q^2,{\b p}_m,{\b p}_1,\tilde{s_1},s_1)
 \dfrac{1}{2M_N}
 \int \dfrac{d^4 k_2}{i(2\pi)^4}
\frac{f^{NN}(\boldkappa_\perp)}
{k_1'{^2}-M_N^2+i\varepsilon} \times\nonumber\\ &&
 \times\left [ \bar u(\b{k}_1,{{\tilde s_1}})
 \Phi_{D}^{{\cal M}_2}(k_1,k_2)  v(\b{k}_2,s_2)\right]
 \label{single}
\ee
 and one can write
\be
      &&T_{\mu}({\cal M}_2,s_1,s_2) \simeq 
       \dfrac{1}{2M_N}
\sum\limits_{\tilde{{s}_1}}J_\nu^{eN}(Q^2,{\b p}_m,{\b p}_1,\tilde{s_1},s_1)
\times  \nonumber\\
&\times&  \left \{\phantom{\frac{d^4}{\pi^4}}\!\!\!\!\!\!\!\!
\left[ \bar u (\b{k}_1,\tilde{s_1}) \Phi_{D}^{{\cal M}_2}(k_1,k_2)
 (\hat k_2+M_N) v(\b{k}_2,s_2)\right ]+ \right. \nonumber\\
 &+& \left. \int \dfrac{d^4 k_2}{i(2\pi)^4}
\frac{f^{NN}(\boldkappa_\perp)}
{k_1'{^2}-M_N^2+i\varepsilon}
 \left[ \bar u(\b{k}_1,{{\tilde s_1}})
 \Phi_{D}^{{\cal M}_2}(k_1,k_2) v(\b{k}_2,s_2)\right] \right \}.
 \label{ttota}
 \ee

It can be seen   that $T_{\mu}^{(0)}$ and $T_{\mu}^{(1)}$
 in Eq. (\ref{ttota})
have   very similar structures, except
 that in $T_\mu^{(1)}$ the vector  $\b{k}_2$ is now an  integration variable, since
 $\b{k}_2 \neq  \b{p}_2$.
  When Eq. (\ref{tensor1}) is evaluated, with $T_\mu$ given by
  Eq. (\ref{ttota}) and  assuming soft $NN$ rescattering
  ( low values of  ${\boldkappa}_\perp$),
  the main contribution  in  the integral
 over $\b{k}_2$ results  from  the region where
  $\b{k}_2\sim\b{p}_2$ and this, in turn,
   originates again a delta function $\delta _{\tilde s_1 s_1}$ (See Appendix \ref{app:fact})
    and  the hadronic tensor  becomes
\be
L_{\mu \nu}^D&=&
\frac{1}{2M_D}\frac13\sum_{{\cal M}_2, s_1,s_2}T_\mu^{\dagger}
({\cal M}_2,s_1,s_2)\cdot T_\nu({\cal M}_2,s_1,s_2)
(2\pi)^4\delta^{(4)}\left( P_D+q-p_1-p_2\right)\simeq\nonumber\\
&\simeq
 &\frac{1}{2}\sum\limits_{{\tilde s}_2,s_1}  \left[ J_\mu^{{eN}\dagger}(Q^2,{\b p}_m,
 {\b p}_1,\tilde{s_2},s_1)\cdot
 J_\nu^{eN}(Q^2,{\b p}_m,{\b p}_1,\tilde{s_2},s_1)\right] \times \nonumber \\
  &\times& \dfrac{1}{(2M_N)^2}\sum\limits_{{{\cal M}_2 \tilde s}_1,s_2}
  \left | \phantom{\frac{f^N}{k^2}}\!\!\!\! \!\!\!\!\!
  \left[\bar u (\b{k}_2,\tilde{s_1})  \Phi_{D}^{{\cal M}_2}(k_1,k_2)
 (\hat k_2+M_N) v({\bf k}_2,s_2)\right]_{k_2=p_2} +
  \right.\nonumber \\
     &+& \left.
    \int \dfrac{d^4 k_2}{i(2\pi)^4}
\frac{f^{NN}(\boldkappa_\perp)}{k_1'{^2}-M_N^2+i\varepsilon}
 [ \bar u(\b{k}_2,{{\tilde s_1}}) \Phi_{D}^{{\cal M}_2}(k_1,k_2)
  v(\b{k}_2,s_2)]\right|^2 \times \nonumber\\
  &\times&  (2\pi)^4\delta^{(4)}\left( P_D+q-p_1-p_2\right)
\label{deutfin1}
\ee

 \noindent and the factorization of the e.m. and the nuclear parts is recovered.
Eq.~(\ref{deutfin1}) could be written in a more familiar form if one integrates
over $k_{20}$ by taking into account the pole in the amplitude
$\Phi_{D}^{{\cal M}_2}(k_1,k_2)$ ($k_{20}=E_{\bf k}$) and  neglecting the
pole from the active propagator, which is located at large values of $k_{20}$
and does not contribute to the integral.
Using (\ref{momdis})-(\ref{vertex}) one obtains ($\int d^4 k/[i(2\pi)^4] \to
\int d^3 k/[(2 E_{\b k}(2\pi)^3]$)
\be
&& L_{\mu\nu}^D
\simeq 2M_D\left[ 2E_{\b {p}_1} 2E_{\b {p}_m}
L_{\mu\nu}^N(Q^2,{\bf p}_m,{\bf p}_1)\right]\times\nonumber\\
&\times&\sum\limits_{\CM_2, s_1,s_2}
  \left |
 \left \langle s_1,s_2|\Psi_D^{\CM_2}({\bf k}_2)\right \rangle_{k_2=p_2}
 + \int \dfrac{d^3 k_2}{2E_{{\bf k}_2}(2\pi)^3}
\frac{f^{NN}(\boldkappa_\perp)}{k_1'{^2}-M_N^2+i\varepsilon}
\left \langle s_1,s_2|\Psi_D^{\CM_2}({\bf k}_2)\right \rangle \right |^2
\times\nonumber\\
&\times&(2\pi)^4\delta^{(4)}\left( P_D+q-p_1-p_2\right).
\label{deutfin}
\ee
By placing the above equation in Eq. (\ref{cross1}), one obtains
 \be
\frac{d^6\sigma}{d E' d\Omega' d{\b p}_m}=  K({ Q}^2,x,{\bf p}_m\,)
 \sigma^{eN}({\bar Q}^2,{\bf p}_m)
 {n}_D^{FSI}({\b{p}}_m))
 \delta (q_0 + M_D - E_{{\b{p}}_1 + {\b{q}}}-E_{{\b{p}}_2}),
 \label{eq19FSI}
 \ee

\noindent where  the  $\it Distorted\,\,\,\, Momentum \,\,\,\,Distribution$ $n_D^{FSI}$ is
\be
&& {n}_D^{FSI}({\b{p}}_m) =\nonumber\\&&
\frac{1}{3}\sum\limits_{\CM_2, s_1,s_2}
  \left |
 \langle s_1,s_2|\Psi_D^{\CM_2}({\bf k}_2) \rangle_{k_2=p_2}
+\int \dfrac{d^3 k_2}{2E_{{\bf k}_2}(2\pi)^3}
\frac{f^{NN}(\boldkappa_\perp)}{k_1'{^2}-M_N^2+i\varepsilon}
\langle s_1,s_2|\Psi_D^{\CM_2}({\b p }_m)\rangle \right |^2=\nonumber\\
&=&\frac{1}{3}\sum\limits_{\CM_2, s_1,s_2}
  \left | {\cal T}_D^{(0)}(\CM_2, s_1,s_2) + {\cal T}_D^{(1)}(\CM_2, s_1,s_2)\right |^2
\label{crossFSI}
\ee

\noindent    and the quantities
 \be
 {\cal T}_D^{(0)}(\CM_2, s_1,s_2)= \left \langle s_1,s_2|\Psi_D^{\CM_2}
 ({\bf k}_2)\right \rangle
 \label{ti0}
 \ee
\noindent and
 \be
{\cal T}_D^{(1)}(\CM_2, s_1,s_2)=\int \dfrac{d^3 k_2}{2E_{{\bf k}_2}(2\pi)^3}
\frac{f^{NN}(\boldkappa_\perp)}{k_1'{^2}-M_N^2+i\varepsilon}
\left \langle s_1,s_2|\Psi_D^{\CM_2}({\bf p}_2)\right \rangle
\label{ti1}
\ee
\noindent can be called
the {\it reduced (Lorentz index independent) amplitudes};
in the above equations    $\Psi_D^{\CM_2}$  is the deuteron wave function 
 and  the spin wave function refers to the two particles in
the continuum.

To sum up we have shown that:
\begin{enumerate}
\item the cross section which includes FSI factorizes
provided: i)
the spin flip part of the $NN$ scattering amplitude can be disregarded, which
is consistent with the high energies we are considering, and ii)
the momentum  transfer $\boldkappa$ in the $NN$ rescattering is small and transverse,
 so that in the integral (\ref{single}) one has
    $\b{k}_2\sim\b{p}_2$ or, equivalently, ${\b k}_2 \simeq {\b p}_m$; this is a reasonable
    approximation, thanks to the behaviour of the elastic  $NN$ scattering
    amplitude,  which is sharply  peaked in the forward direction;
\item in the eikonal approximation and neglecting
 the spin dependence (spin-flip part)  of the $NN$-amplitude,
 the FSI is not affected by the spin structure of the
 wave functions of the deuteron  and the
two-body final state. This means that
in computing the Feynman diagrams,   the intermediate
spin algebra  can be disregarded,  and  only the scalar part
of the corresponding
vertex functions can be considered, using Eq.~(\ref{vertex}) to define
the scalar parts of the wave functions.
Then the resulting amplitude has to  be merely  sandwiched between the
spin functions of initial and final particles.
\end{enumerate}

These  conclusions can be generalized to a nucleus $A$, for which 
the  cross section of the process
$A(e,e'p)(A-1)$  is given by the following expression:
\be
&&
\frac{d^6\sigma}{dE'd\Omega 'd{\bf p}_m}=
      K({ Q}^2,x,{\bf p}_m)\, \sigma^{eN}({\bar Q}^2,{\bf p}_m)
P_A^{FSI}({\b p}_m, E_m)
\label{crosfsi}
\ee
where $P_A^{FSI}({\bf p}_m,E_m)$ is
the ${\it Distorted\,\,\,\, Spectral\,\,\,\, Function}$
\be
P_A^{FSI}({{\b p}_m},E_m)&=& \frac {1}{(2\pi)^3} \frac{1}{2J_A+1}\,\,\, \sum_{f}\,\,\,
\sum_{{\cal M}_A,\,{\cal M}_{A-1},\,\,s_1}
\left | \sum_{n=0}^{A-1} {\cal T}_A^{(n)}(\CM_A,{\cal M}_{A-1}, s_1) \right |^2\times \nonumber\\
&\times& \delta\left( E_{m}-(E_{A-1}^f + E_{min})\right)
\label{pdistor}
\ee
and $n$ denotes the order of rescattering.  In what follows the
distorted momentum
distributions   for the Deuteron (Eq. (\ref{crossFSI})) and the distorted Spectral Function
 for $^3He$  (Eq. (\ref{pdistor})) will be calculated within the
 GA and GEA.

\subsection{The process $^2H(e,e'p)n$ within  the GA and GEA.}

 Let us now calculate the reduced amplitude ${\cal T}_D^{(1)}$
in the process $^2H(e,e'p)n$,  taking FSI into account by the
 GEA.  This amounts to replace the energy denominator in Eq. (\ref{ti1}) by
 its generalized eikonal approximation. To this end,  we will consider both the
 "canonical" case,  when the value of the 3-momenta transfers $|{\bf q}|$
 is so high
 that ${\bf q}  \simeq {\bf p}_1$, with the $z$-axis naturally directed along ${\bf q}$,
 as well as the case  of smaller values of ${\bf q}$, but high values of
 ${\b p}_1$, when  ${\bf q}$ and
 ${\bf p}_1$ may point to different directions, in which case  the $z$-axis
 is oriented along ${\bf p}_1$.

Remembering that $\kappa = p_1-k_1^{'}= k_2-p_2$, the energy denominator
  can be written as
follows
\be &&
k_1'^2 - M_N^2 =
(p_1-\kappa)^2 - M_N^2 =
-2p_1\kappa + \kappa^2=
2|{\bf p}_1|
\left(\kappa_z +\dfrac{\kappa_0(\kappa_0-2E_{{\bf p}_1})}{2|{\bf p}_1|}
-\dfrac{\boldkappa^2}{2|{\bf p}_1|}    \right )\approx \nonumber\\
&&\approx  2| \b{p}_1| \left( \kappa_z -
\dfrac{E_{{\bf k}_1+{\bf q}} + E_{{\bf p}_1}}{2|{\bf p}_1|} \kappa_0 \right)
\approx
2| \b{p}_1| \left( \kappa_z + \Delta_z\right),
\ee
 where
\be
\Delta_z =   \dfrac{E_{{\bf k}_1+{\bf q}} + E_{{\bf p}_1}}{2|{\bf p}_1|}
\left( E_m  -|E_A|\right)
\label{delta}
\ee
    and  the relation
\be
\kappa_0=   E_{{\bf p}_1}- E_{{\bf k}_1+{\bf q}} \approx
-(E_m-|E_A|)
\label{denomina}
\ee
 resulting from energy conservation
 $q_0+M_D = E_{\b {p}_1} + E_{\b {p}_2}$ has been used.

By changing  the normalization of the $NN$ amplitude
from the covariant one to the non relativistic analogue  ($E_{\b p}\simeq
M_N$), one has

\be&&
\dfrac{f^{NN}(\boldkappa_\perp)}{4E_{{\b p}}|{\b{p}}_1|}
\approx
\dfrac{f^{NN}(\boldkappa_\perp)}{4M_N|{\b{p}}_1|} = a^{NR}(\boldkappa_\perp)
\equiv i\int d^2\b{b} {\rm e}^{i\boldkappa_\perp\b{b}}\ \Gamma(\b{b})
\ee
\noindent and ${\cal T}_D^{(1)}$ becomes
\be
{\cal T}_D^{(1)} ({\cal M}_2,s_1,s_2)=
\int \dfrac{d^3 k}{(2\pi)^3}
a^{NR}(\boldkappa_\perp)
\dfrac{1}{\kappa_z+\Delta_z + i\varepsilon}
\left\langle s_1,s_2 | \Psi_D^{{\cal M}_2}(\b{p} )
\right\rangle .
\ee
\noindent Using
\be
\frac{1}{\kappa_z+\Delta_z + i\epsilon} =  -i\int \theta(z)\,\, {\rm e}^
{i(\kappa_z+\Delta_z)\cdot z}dz
\label{cospace}
\ee

\noindent we obtain, in coordinate space,
\be
{\cal T}_D^{(0)} ({\cal M}_2,s_1,s_2)+{\cal T}_D^{(1)} ({\cal M}_2,s_1,s_2)=
\left\langle \
s_1,s_2  \left( 1-  \theta(z) {\rm e}^{i\Delta_z z}\Gamma(\b{b}) \right)
{\rm e}^{-i\b{p}_m\b {r}}
|\Psi_D^{{\cal M}_2}(\b{r})
\right \rangle .
\label{matrixdis}
\ee

\noindent
As a result, the cross section will read as follows
\be
&&
\frac{d^6\sigma}{dE' d\Omega' d{\b p}_m}=
K({Q}^2,x,{\bf p}_m)\,
 \sigma^{eN}({\bar Q}^2,{\bf p}_m)
  n_D^{FSI}(\b{p}_m)
 \delta (M_D+\nu-E_{\b{p}_1}-E_{\b{p}_m})
 \label{eq191}
\ee

\noindent with the   {\it distorted momentum distributions} $n_D^{FSI}$   defined by
\be
 n_D^{FSI}( {\bf p}_m) =
\frac13\frac{1}{(2\pi)^3} \sum\limits_{{\cal
M}_2,S_{23}} \left | \int\, d  {\bf r} \chi_{S_{23}}^\dagger\Psi_{D}^{{\cal
M}_2 \dagger}( {\bf r}) {\cal S}_{\Delta}^{FSI}( {\bf r}) \,\exp (-i
{\bf p}_m {\bf r}) \right |^2, \label{ddistr}
\ee

\noindent where   $S_{\Delta}^{FSI}( {\bf r})$, which
describes  the  final state interaction
between the hit nucleon and the spectator, is
\be
{\cal S}_{\Delta}^{FSI}({\bf r})=  1-  \theta(z){\rm e}^{i\Delta_z z}\Gamma(\b{b})
\label{esse}
\ee

\noindent with $\b{r}=(\b{b},z)$.
In the above formulae  the z-axis is along  ${\b{p}}_1$;
it should be pointed out, however,  that
  at large values of the momentum transfer,
  the hit nucleon propagates almost along $\b q$ so that by choosing the $z$-axis
  along the three-momentum transfer and neglecting the virtuality
  of the struck nucleon before and after interaction, one can write \cite{strikman}
  \be
k_1'^2 - M_N^2=
(k_1+ q)^2 - M_N^2 \approx
2|{\bf q}|   \left( \kappa_z + \Delta_z\right),
\label{denmark}
\ee

where
\be
\Delta_z = \frac{q_0}{|{\b q}|} E_m.
\label{deltamark}
\ee

 It can be seen that the FSI factor (\ref{esse})
in the GEA differs from the one of the standard GA
\cite{CKT,niko1, niko}, simply by  the additional factor  ${\rm e}^{i\Delta_z z}$.
 It should be pointed out that whereas the well
  known factor $\theta(z)$  \cite{niko1,niko}
  originates from the non relativistic reduction of the covariant  Feynman diagrams and guarantees
 the correct time ordering of the rescattering processes,   the quantity
 $\Delta_z$ is of a  pure nuclear structure origin and, as it can be seen
 from Eq. (\ref{matrixdis}),  represents a
 correction to the parallel component of the missing momentum. Therefore
 the corrections from $\Delta_z$ are expected to  be important in  parallel
 kinematics at $|\b{p}_z|\simeq\Delta_z$.
 As we
 shall see from the results of our  calculations  performed  in perpendicular
 kinematics
 in the range  $|{\b p}_m|\leq 600\,MeV/c$ and  $E_m \leq 100 \, MeV$,  one always has
 $| \Delta_z|\ll|\b {p}|_\perp$ with  ${q_0}/{|{\b q}|}\simeq 1$, so that
  $\Delta_z $ is always very small.  We can therefore anticipate that  effects of
   $\Delta_z $
   on
   the experimental
  data we have considered is also very small.

\subsection{The processes $^3H(e,e'p)^2H$ and $^3H(e,e'p)(np)$
 within the  GA and GEA.}

Let us now consider the three-body system. The distorted Spectral
 Function is given by  Eq. (\ref{pdistor})

\be
P_{He}^{FSI}({{\b p}_m},E_m)&=& \frac {1}{(2\pi)^3} \frac{1}{2}\,\,\, \sum_{f}\,\,\,
\sum_{{\cal M}_3,\,{\cal M}_{2},\,\,s_1}
\left | \sum_{n=0}^{2} {\cal T}_A^{(n)}(\CM_3,{\cal M}_{2}, s_1) \right |^2\times \nonumber\\
&\times& \delta\left( E_{m}-(E_{2}^f + E_{min})\right),
\label{pdistor3}
\ee

\noindent where the magnetic quantum number ${\cal M}_2$ refers
either to the deuteron or to
 the two nucleon in the continuum,
depending upon the break-up channel we are considering ($E_{min}=E_3-E_2 \,(E_{min}=E_3)$
 for the
two-body (three-body) break-up channel).
The diagrams representing the
rescattering processes are shown in Fig. 6. The evaluation
 of these diagrams follows
the standard procedure adopted for the deuteron.
Let us illustrate it in the case of the   3bbu considering,
 for ease of presentation,
the single scattering diagram
of Fig. 6 b).  After
integration over $k_{20}$ and $k_{30}$ in the corresponding poles
of the propagators of the spectators ($k_{20}=E_{{\bf k}_2}$
and $k_{30}=E_{{\bf k}_3}$),  we  obtain

 \be
 &&{\cal T}_3^{(1)}({\cal M}_3,s_1,s_2,s_3) =
 \int\frac{d^3 k_2}{2E_{{\bf k}_2}(2\pi)^3} \frac{d^3 k_3}{2E_{{\bf k}_3}(2\pi)^3}
\times\nonumber \\
 &\times&
  \frac{G_{He\to 1(23)}(k_1,k_2,k_3,s_1,s_2,s_3)}{(k_1^2-M_N^2)}
\frac{f_{NN}(p_1-k_1')}{k_1'^2-M_N^2 }
 \frac{
 G^+_{(23)\to f }(k_2',k_3,s_2,s_3)}{(k_2'^2-M_N^2)},
 \label{eq15}
 \ee

\noindent where the overlaps of the vertex functions $G_i$ are
\be
&&
 G_{He\to 1(23)}
 (k_1,k_2,k_3, s_1,s_2,s_3)
 =\langle {\bf k}_1,s_1, {\bf k}_2,s_2, {\bf k}_3,s_3|
  G_{He\to 1(23)}(\CM_3,{\bf P}_3)\rangle;
\\ &&
  G_{(23)\to f}(k_2,k_3,s_2,s_3)
= \langle  {\bf k}_2,s_2, {\bf k}_3,s_3|
  G_{(23)\to f }
  (\CM_{23},S_{23},{\bf P}_2,E_2^f)\rangle;
  \label{amhe}
\ee

  The vertex functions $G_i$ are
  replaced by the non relativistic overlap functions
 according to the general convention (we omit for ease of presentation the proper
 normalization factors)

 \be
  \langle s_1,s_2,s_3 | \Psi_{He}^{{\cal M}_3}(\b {k}_1,\b {k}_2,\b {k}_3)\rangle
   \approx  \frac{ G_{He\to 1(23)}(k_1,k_2,k_3, s_1,s_2,s_3)}{(k_1^2-M_N^2)}
  \ee
  \noindent and, using the completeness relation when summing  over $s_2$ and $s_3$, one gets

 \be
 &&{{\cal T}_3^{(1)}}(Q^2,s_1,S_{23}) =
 \int\frac{d^3 k_2}{2E_{{\bf k}_2}(2\pi)^3}
 \frac{d^3 k_3}{2E_{{\bf k}_3}(2\pi)^3} \times\nonumber\\
 &\times&\Psi_{(23)}^f({\bf k}_3,{\bf k}_2'; S_{23})
  \frac{f_{NN}(\boldkappa)}{({k_1'}^2 - M_N^2+i\epsilon)}
 \langle  s_1|\Psi_{He}^{{\cal M}_3}(\b {k}_1,\b {k}_2,\b {k}_3)\rangle .
 \ee

Following the procedure adopted for the deuteron,  we obtain

\be
&&{{\cal T}_3^{(1)}}(Q^2,s_1,S_{23})=\nonumber\\
&=&\int\frac{d^3\kappa}{(2\pi)^3\,2E_{{\b k}_2}}
  \Psi_{(23)}^f({\bf k}_3,{\bf k}_2'; S_{23})
  \frac{f_{NN}(\boldkappa)}
 {k_1'^{2}-M_N^2+i\varepsilon}
 \langle s_1|\Psi_{He}^{{\cal M}_3}(\b {k}_1,\b {k}_2, \b {k}_3)\rangle \approx\nonumber\\
 &\approx&
   \int\frac{d^3\kappa}{(2\pi)^3}
  \Psi_{(23)}^f({\bf k}_3,{\bf k}_2'; S_{23})
  \frac{f_{NN}(\boldkappa)/4M_N|\bf {p}_1|}
 {
\left(  \kappa_z+\Delta_z+i\epsilon\right)}
   \langle s_1|\Psi_{He}^{{\cal M}_3}(\b {k}_1,\b {k}_2, \b {k}_3)\rangle ,
 \label{eq16}
 \ee

 \noindent where
 \be
\Delta_z =  \dfrac{E_{{\bf k}_1+{\bf q}} + E_{{\bf p}_1}}{2|{\bf p}_1|}
 (E_m -E_3).
\label{delta3}
\ee

\noindent Including  also the 2buu channel, we can write,
 in  coordinate space,
 \be
 P_{He}^{FSI}({\b p}_m,E_m) =  P_{gr}^{FSI}({\b p}_m,E_m) +
  P_{ex}^{FSI}({\b p}_m,E_m),
\label{pitotfsi}
\ee
\noindent where
\be
P_{gr}^{FSI}({\b p}_m,E_m) = n_{gr}^{FSI}({\b p}_m)\delta(E_m - (E_{3}-E_2))
\label{pgrfsi}
\ee

\noindent with
\be
n_{gr}^{FSI}({\b p}_m)=\frac{1}{(2 \pi)^3} \frac{1}{2}
  \sum_{\CM_3, \CM_2,s_1}
  \left | \int {\rm e}^{i\boldrho{\b p}_m}
 \chi_{\frac12 s_1}^{\dagger} \Psi_{D}^{{\CM_2}\,\dagger}(\b{r} )
  {\cal S}_{\Delta}^{FSI}(\boldrho,{\bf r})
 \Psi_{He}^{\CM_3}(\boldrho,\b{r})   d \boldrho d {\bf r} \right |^2
\label{ngrfsi}
\ee

\noindent and

 \be
 P_{ex}^{FSI}({\b p}_m,E_m)&=& \frac{1}{(2 \pi)^3} \frac{1}{2}
  \sum_{{\cal M}_3, S_{23}, s_1}
       \int \frac{d^3 \b {t}}{(2\pi)^3}
  \left | \int{\rm e}^{i\boldrho{\b p}_m} \chi_{\frac12 s_1}^\dagger
 \Psi_{np}^{\b{t}\dagger}(\b{r}){\cal S}_{\Delta}^{FSI}(\boldrho,\b{r})
  \Psi_{He}^{{\cal M}_3}(\boldrho,\b{r})  d\boldrho d {\bf r}
  \right |^2\times\nonumber\\
&\times&\delta \left( E_m - \frac{\b {t}^2}{M_N} - E_3 \right).
\label{piexfsi}
\ee

\noindent The FSI factor   ${\cal S}_{\Delta}^{FSI}$  describes the single and double
 rescattering of  nucleon "1"
with the spectators "2" and "3", and has the following form

\be
{\cal S}_{\Delta}^{FSI}(\boldrho,\b{r}) ={\cal S}_{(1)}^{FSI}(\boldrho,\b{r})+
{\cal S}_{(2)}^{FSI}(\boldrho,\b{r})
\label{totalS}
\ee

\noindent with  the single scattering contribution   ${\cal S}_{(1)}^{FSI}$
 given by
\be
{\cal S}_{(1)}^{FSI}(\boldrho,\b{r})=1-\sum\limits_{i=2}^3
\theta(z_i-z_1){\rm e}^{i\Delta_z (z_i-z_1)} \Gamma (\b {b}_1-\b{b}_i)
\label{essesingle}
\ee

\noindent and the
  double scattering contribution by  (Ref. \cite{strikman,marknew}).
 \be &&
{\cal S}_{(2)}^{FSI}(\boldrho,\b{r}) =
\left[\theta(z_2-z_1)\theta(z_3-z_2){\rm e}^{-i\Delta_3(z_2-z_1)}
{\rm e}^{-i(\Delta_3-\Delta_z)(z_3-z_1)}+   \right. \nonumber \\&+&
     \left.
\theta(z_3-z_1)\theta(z_2-z_3){\rm e}^{-i\Delta_2(z_3-z_1)}
{\rm e}^{-i(\Delta_2-\Delta_z)(z_2-z_1)}\right]  \times
 \Gamma(\b {b}_1-\b{b}_2)\Gamma(\b {b}_1-\b{b}_3),
  \label{essedouble}
 \ee

 \noindent where  $\Delta_i=(q_0/|{\b q}|)(E_{{\b p}_i} - E_{{\b k}_i^{'}})$ and $\Delta_z$
 is given by Eq. (\ref{deltamark}).

 When $\Delta_z=0$, the  familiar form   for   ${\cal S}^{FSI}$ is obtained, namely
\begin{equation}
{\cal S}^{FSI}(\boldrho,\b{r}) =
\prod\limits_{i=2}^{3}\ \left[ 1-\theta(z_i-z_1)\,
\Gamma({\bf b}_i-{\bf b}_1)\right ],
\label{eq20}
\end{equation}
 and when  $\Gamma = 0$,
 the distorted Spectral Function (\ref{pitotfsi}) transforms into
  the usual Spectral Function  (\ref{eq13}).

Using Eq. (\ref{pitotfsi}), the cross section of the process $^3He(e,e'p)X$
 ($X=D$ or $(np)$) assumes the following form
\be
&&
\frac{d^6\sigma}{dE' d\Omega' d{\b p}_m}=
K({Q}^2,x,{\bf p}_m)\,
 \sigma^{eN}({\bar Q}^2,{\bf p}_m)
  P_{He}^{FSI}(\b{p}_m,E_m).
 \label{crosshe}
\ee

\section{Results of the calculations}\label{serR}
We have used Eqs. (\ref{eq191}), (\ref{esse}), (\ref{crosshe}) and (\ref{essedouble})
to calculate the cross sections of the processes  $^2He(e,e'p)n$,
$^3He(e,e'p)^2H$ and $^3He(e,e'p)(np)$. All
calculations  have been performed
using the following well known parametrization of the
profile function $\Gamma({\bf b})$
\be
\Gamma({\bf b}) =\frac{\sigma_{NN}^{tot}(1-i\alpha_{NN})}{4\pi b_0^2}
{\rm e}^{-{\bf b}^2/2b_0^2},
\label{profilo}
 \ee
 where $\sigma_{NN}^{tot}$ is the total $NN$ cross section,
 $\alpha_{NN}$ the ratio of the real to imaginary part of the forward $NN$ amplitude,
 and  $b_0$ the slope of
  the differential elastic $NN$ cross section. The values of the energy dependent
  quantities
    $\sigma_{NN}^{tot}$ and $\alpha_{NN}$ have been taken from Ref. \cite{said}.
    For the electron-nucleon cross section $\sigma^{eN}({\bar Q}^2,{\bf p}_m)$
we used the De Forest  $\sigma_{cc1}^{eN}({\bar Q}^2,{\bf p}_m)$
cross section  \cite{forest}. All two- and three-body wave functions are direct
solutions of the  non relativistic Schr\"odinger equation, therefore
our calculations  are fully parameter free.

    Calculations have been performed in PWIA and including the full rescattering within
    the  GA and  GEA  by evaluating
    the Feynman diagrams shown in Figs.~5 and 6.
    It should be pointed out that, apart from minor differences (e.g. the structure of
    $\Delta_z$ for complex nuclei) which do not affect the numerical results, our GEA
    is essentially the same as the one developed in \cite{strikman,marknew}.
    \subsection{The process $^2H(e,e'p)n$}
Our results for  the  process $^2H(e,e'p)n$
are  compared  in  Figs. \ref{fig7}, \ref{fig8} and \ref{fig9}
with three different sets of experimental data,
 covering different kinematical ranges,  namely the experimental
 data from NIKHEF \cite{nikhef}, SLAC \cite{bulten}, and Jlab \cite{ulmer}.
The relevant kinematical variables in the three experiments are as follows: i)
$0.1 \leq Q^2 \leq 0.3$, $0.3 \leq x \leq 0.6$ \cite{nikhef}; ii) $1.2 \leq Q^2 \leq 6.8$, $ x \simeq 1$
\cite{bulten}; iii) $Q^2\simeq 0.665\,\,\, (GeV/c)^2$ , $|{\b q}|
\simeq 0.7\,\,\,
GeV/c$, $x\simeq 0.96$ \cite{ulmer}.
In Figs. \ref{fig7} and \ref{fig9} the  theoretical cross section corresponding to
 Eq.
(\ref{eq191}), namely
\be
\frac{d^5\sigma}{dE' d\Omega' d\Omega_{{\b p}_m}}=
f_{rec}\,\,{\cal K}(Q^2,x, {\b p}_m)\,
 \sigma_{cc1}^{eN}({\bar Q}^2,{\bf p}_m)
  n_D^{FSI}(\b{p}_m)
 \label{crossD}
\ee

\noindent is compared with the corresponding
data, whereas
in Fig. \ref{fig8} we compare, as in Ref. \cite{ulmer},
 the effective momentum
distributions   $N_{eff}(p_m)$  (or reduced cross section)
 defined by \cite{ulmer}
\be
N_{eff}(|{\b p}_m|)=
\frac{d^5\sigma^{exp}}{d\Omega'dE'd\Omega_{{\b p}_m}}
\left [f_{rec}\,\,{\cal K}\,\, \sigma_{cc1}^{eN} \right]^{-1},
\label{neff}
\ee
where in Eqs. (\ref{crossD}) and (\ref{neff})  $f_{rec}$ and   ${\cal K}$
are kinematical factors which arise from the integration over $dT_{{\b p}_1}$.

The results presented in Figs. 7, 8 and 9 exhibit
 a general satisfactory
agreement between theoretical calculations and experimental data,
particularly in view of  the wide
range of kinematics covered by the data we have considered. Figs. 8 and 9
show however that quantitative disagreements with data exist in some regions.
Particularly
 worth
being noted is  the disagreement in the region around
 $|{\b p}_m|\simeq 0.25\,\, GeV/c$
appearing in Fig. 8. We did not try to remove such a  disagreement
 by adjusting the quantities entering the profile function  (\ref{eq191}),
 but it turns out that
$n_D^{FSI}$ in  the
region around  $|{\b p}_m|\simeq\,\,\, 0.25 \,\,\,GeV/c$ is rather sensitive to
 the
value of $\alpha_{NN}$. 
The NIKHEF kinematics deserves a particular comment. As a matter of fact, the four
momentum transfer in this experiment is rather low and one might rise doubts as to the validity
of the eikonal approximation. In this respect, it
should however be pointed out, that
 what really matters in GA and GEA is the relative three-momentum
of the hit  nucleon  with respect to the $A-1$ system;
in the NIKHEF experiment, due to the large value of the energy transfer,
the three momentum transfer  is also large, and $\gamma^*$ absorption occurs on
a  proton moving
along  $\bf q$,  with the recoiling neutron moving with low momentum against
$\bf q$; the resulting proton-neutron
relative momentum is of the order of few hundreds $MeV/c$, which, though representing
 the
lower  limit for the validity of the
eikonal approximation, still appears, according to our results,  to be suitable for the application of the
GEA: as a matter of fact  our results appear to be in reasonable agreement
 with the ones obtained within approaches
which are better justified at low energies, like, e.g., the ones  presented,
in Refs. \cite {hart1,hart2,hart3,laget1,tjon2}.

Let us conclude this Section by stressing that, as far as the effects of MEC and $\Delta$ isobar
excitations are concerned, these have been
found to be very
small ($\simeq 5-6 \%$)
in the SLAC kinematics ( see \cite{bulten}), with
the results from  \cite{laget4}  exhibiting  the same trend also in the
Jlab kinematics.
Eventually, we should remark that
the results of the GA and GEA differ by only few percent and cannot
be distinguished in the Figures.

\subsection{The processes $^3He(e,e'p)^2H$ and $^3He(e,e'p)(np).$}
 Calculations for the three-body systems are very involved, mainly because of
 the complex structure of the  wave function of Ref.~\cite{pisa}, which
  is given  in a mixed
   $\left(L_\rho,X,j_{23},S_{23}\right)$ representation, including
   angular
   momentum values  up to  $L_\rho=7$ and $j_{23}=8$
   (a total of 58 configurations with
   different combinations of
   $\left(L_\rho,X,j_{23},S_{23}\right)$ quantum numbers).
  Correspondingly, the wave function of the spectators
  (the deuteron  or the continuum two-nucleon  states) is given
  in a
  $JLS$-scheme (see Appendix~\ref{app}).
  We would  like to stress, that no approximations have
  been made in the evaluation
  of the single and double scattering contributions to the FSI: proper
  intrinsic coordinates have been used and the energy dependence of the profile
  function  has been taken into account in the properly chosen CM system of the
  interacting pair.   The Feynman diagrams which have to be  evaluated,
  both for the 2bbu and 3bbu channels are shown in Fig.~6.

      \subsubsection{The two-body break-up  channel $^3He(e,e'p)^2H$.}

The 2bbu channel cross section
\be
&&
\frac{d^5\sigma}{dE' d\Omega' d{\Omega}_{{\bf p}_1}}=
 K_{2bbu}({ Q}^2,x,{\bf p}_m)\,
 \sigma_{cc1}^{eN}({\bar Q}^2,{\bf p}_m)
 n_{gr}^{FSI}({\b p}_m)
 \label{crossgr}
\ee
obtained from Eq. (\ref{crosshe}), with    $n_{gr}^{FSI}({\b p}_m)$
 given by Eq. (\ref{ngrfsi}),
 is compared   in Fig. 10  with recent
experimental data from  Jlab Collaboration \cite{jlab1}.
 The relevant kinematical variables in the experiment are
 $|\b{q}|=1.5\,\, GeV/c$, $q_0=0.84\,\, GeV$, $Q^2=1.55 \, (GeV/c)^2$,
 and  $x\approx 1$.
 The cross section is presented as a function of the missing momentum
  $|{\b p}_m|$
 (which, for the  $^3He(e,e'p)D$-process,
 exactly coincides with the final deuteron momentum).
In PWIA  the cross section is directly proportional to  $n_{gr}$ (Eq. (\ref{ngr}))
shown in the left panel of Fig.~\ref{fig3}.
 It  can be seen
 that  up to
 $|{\b p}_m|\sim 400\,\, MeV/c$,  the PWIA and FSI results are almost the same and fairly
 well agree with the experimental data, which means, in turn, that
 the 2bbu    $^3He(e,e'p)^2H$ does provide information on $n_{gr}$;
 on the contrary,
 at larger values of $|{\b p}_m|\geq 400\,\, MeV/c$  the PWIA
appreciably underestimates the experimental data. It is very
gratifying to see that   when FSI is taken into account,  the disagreement
 is fully removed and an overall very good agreement
between theoretical predictions and experimental data is obtained.
 It should be pointed out  that
 the experimental data shown in Fig. \ref{fig10}
 correspond to the  perpendicular kinematics,
 when
 the deuteron momentum (the missing momentum) is always
 almost perpendicular to the momentum transfer $\b q$; in such
a kinematics the effects from  FSI are maximized, whereas in the so called
parallel kinematics, they are minimized
 (see, e.g. ~\cite{mark}, ~\cite{niko}, ~\cite{mor01}). The kinematics
 therefore reflects itself in the relevance of the
 calculated FSI;  as a matter of fact, we have
 found that the effects of the FSI calculated
 either within  the  GA  or GEA approximations,
 differ only by a few percent, which was
 expected in view of the observation
 that the factor ${\Delta_z}$ (Eqs. (\ref{delta3}) or (\ref{deltamark})) affects only
 the longitudinal component of ${\b p}_m$ and therefore has minor effects on
  the data we have  considered.
  The effects of MEC and  $\Delta$ isobar contributions have been
  estimated in
  \cite{laget4}
   and found negligible up to about $p_m \simeq 600 MeV/c$.

    \subsubsection{The three-body break-up  channel $^3He(e,e'p)(np)$.}
From Eq. (\ref{crosshe}), we obtain the  cross section for the 3bbu in the
following form
 \be
&&
\frac{d^6\sigma}{dE' d\Omega' d{\Omega}_{{\bf p}_1}dE_m}=
K_{3bbu}({ Q}^2,x,{\bf p}_m)\,
 \sigma_{cc1}^{eN}({\bar Q}^2,{\bf p}_m)
 P_{ex}^{FSI}({\b p}_m, E_m),
 \label{crossexp}
\ee
\noindent where  $P_{ex}^{FSI}({\b p}_m, E_m)$ is given by Eq. (\ref{piexfsi}).
 We have calculated Eq. (\ref{crossexp}) in correspondence of
 two different kinematical ranges: the one from Ref. \cite{saclay} and the
 one corresponding to  the
experimental data from Jlab \cite{jlab2}.
 Contrary to the 2bbu channel,  the 3bbu cross section depends
  upon an extra kinematical
  variable, the removal energy $E_m$,  and corresponds to the process in
  which three particles interact in the continuum.
  We have  considered three different theoretical approaches, namely:

  \begin{enumerate}
  \item the Plane Wave Approximation (PWA), when  FSI effects are completely ignored
  , i. e.
  the three particles in the continuum are described by plane waves;
  \item the  Plane Wave Impulse Approximation (PWIA),
  in which the struck nucleon is described in the continuum
  by a plane wave and the spectator pair  is
  described by the continuum solution of the Schr\"odinger equation (obviously, in the case of the
  deuteron the PWIA coincides with the PWA);
  \item the full FSI,  when the struck nucleon interacts in the continuum
  with the nucleons of the spectator pair via the standard GA or the more refined
  GEA.
  \end{enumerate}

    In Fig. \ref{fig11} the results of our calculations are compared with
    the experimental data from Ref. \cite{saclay}.
    In the experiment, which corresponds  to a
  relatively low beam energy ($E = 0.560\ GeV$),
  the scattering angle  ($\theta_e= 25^o$) and the energy transfer
  ($q_0 = \ 0.32 GeV$) were kept constant,  and protons with different
  values of the missing momentum and energy were detected in correspondence
  of several values of the proton emission angle
   $\theta_{{\b p}_1}$, {
  \it viz}\,\, $\theta_{{\b p}_1} =  45^o,\, 60^o, \,
   90.5^o,\, 112^o$ and $142.5^o$. The
   kinematics is far from the quasi elastic peak
   ($ x\simeq 0.1$) and the values of the four- and three-momentum transfers
   are low  ($Q^2\simeq 0.03\ (GeV/c)^2$ and $|\b{q}|=0.28 \ GeV/c$).
    At first glance this would invalidate the use of the eikonal approximation;
     however,
    a detailed analysis of  the kinematics, shows that the value of
     both ${\b p}_1$  and ${\b p}_m$  are rather large ($400 - 600\ MeV/c$),
   and so is the value of the angle  between them
    ($\theta_{\widehat {\b{p_1}\b{p}}_m}\sim 150 ^o$); thus  the
    momentum of the struck nucleon relative to the spectator pair
    is high enough to make the use of the eikonal approximation justified.
   Moreover,
    the values of the  experimentally measured
   missing momenta and missing energy at each value of
   $\theta_{\b{p_1}}$,
   always cover the kinematical range where the
 condition for two nucleon correlation mechanism
   $E_m\sim \displaystyle\frac{\b {p}_m^2}{4M_N}$  holds; as a matter of fact, as
   it can be seen from Fig. 11,
   the position of the bumps in the cross section are reasonably predicted by the
   PWA and PWIA.

    The results presented in Fig.~\ref{fig11} clearly show  that
    with increasing missing momentum,  the experimental peak
    moves to higher values of missing energy, in qualitative agreement with
    the two-nucleon correlation mechanism. More important, it can be seen that
    at the highest value of $|{\b p}_m|$ ($\theta_{{\b p}_1} =112^o$) the effects
    of FSI,  both in the spectator pair and between the struck nucleon and
    the spectator
     pair, is very small. The reason for such a behaviour is as follows:
   the kinematics  of the experiment  is not purely perpendicular: the relation
   between $|\b{p}_{m\perp}|$ and $|\b{p}_m|$  is such that
   $|\b{p}_{m\perp}|\sim \displaystyle\frac12 |\b{p}_m|$, so that the
    dominant role
  played by FSI in the
   purely perpendicular kinematics is decreased with increased values of
   $\theta_{{\b p}_1}$.

\begin{figure}[!htp]
\centerline{
      \epsfysize=4.5cm\epsfbox{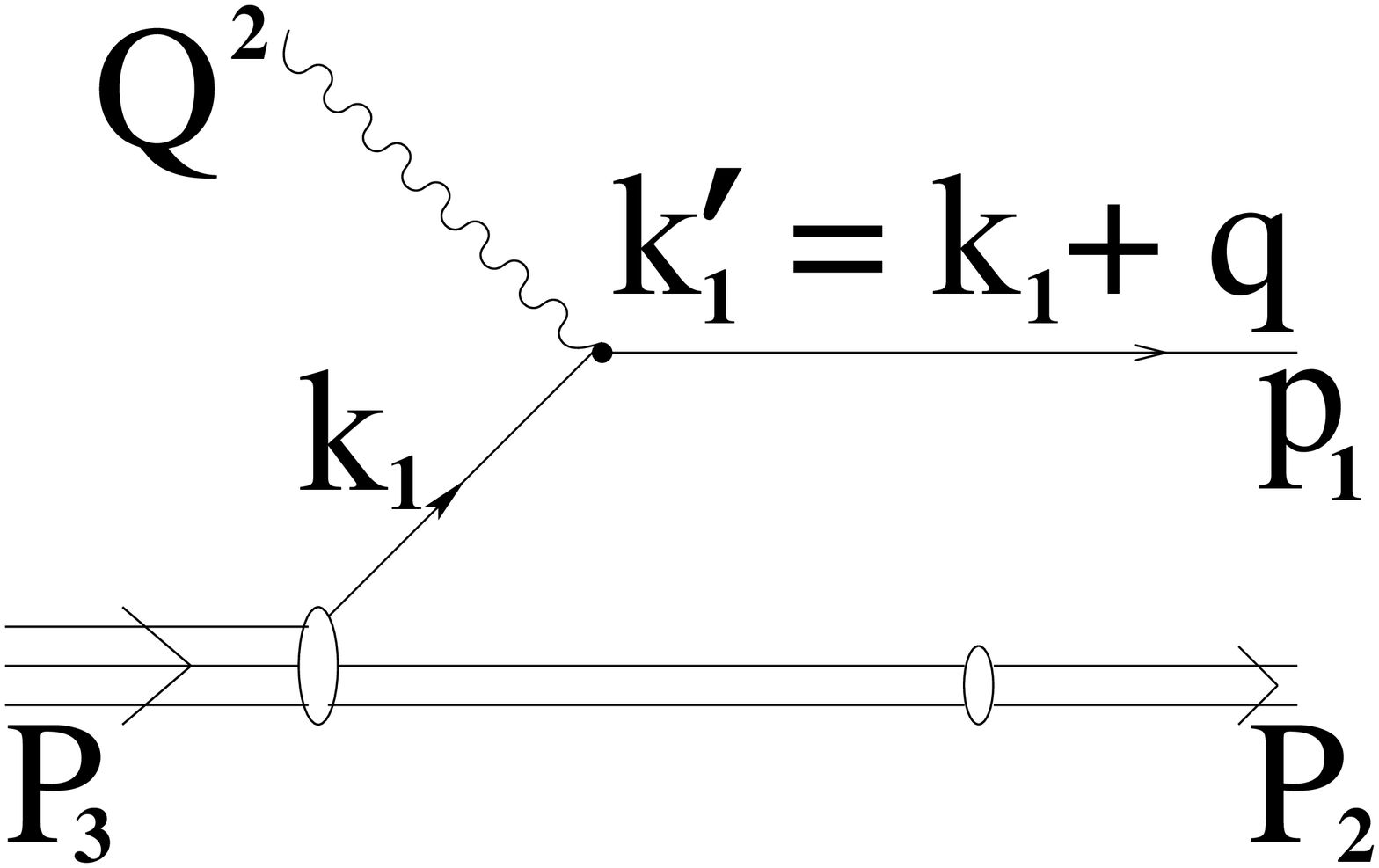}}
      \vspace{.5cm}
\centerline{\large\textbf{a)}\hspace{1cm}\normalsize}
      \vspace{.5cm}
\centerline{
      \epsfysize=4.5cm\epsfbox{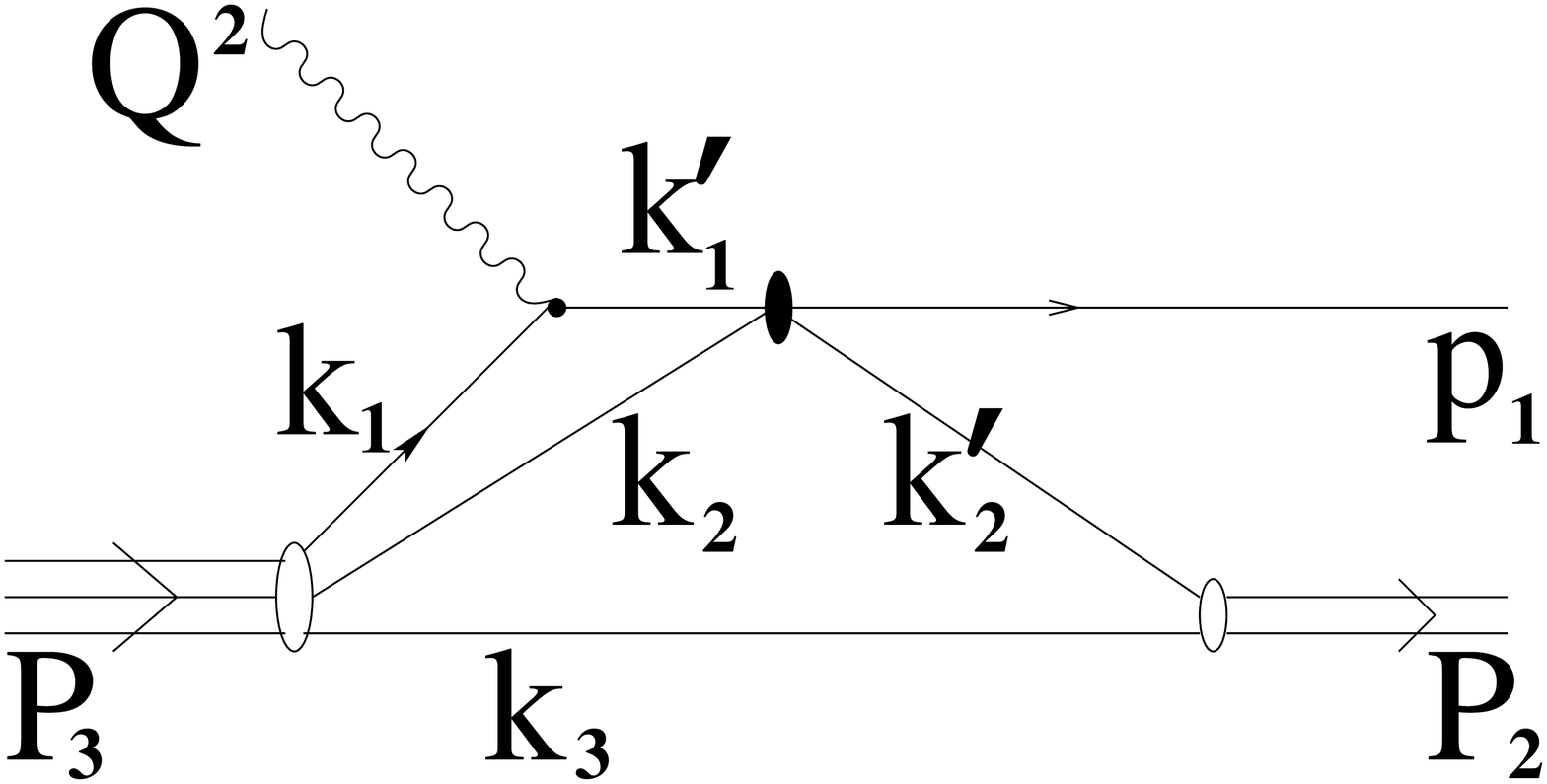}}
      \vspace{.5cm}
\centerline{\large\textbf{b)}\hspace{1cm}\normalsize}
      \vspace{.5cm}
\centerline{
      \epsfysize=4.5cm\epsfbox{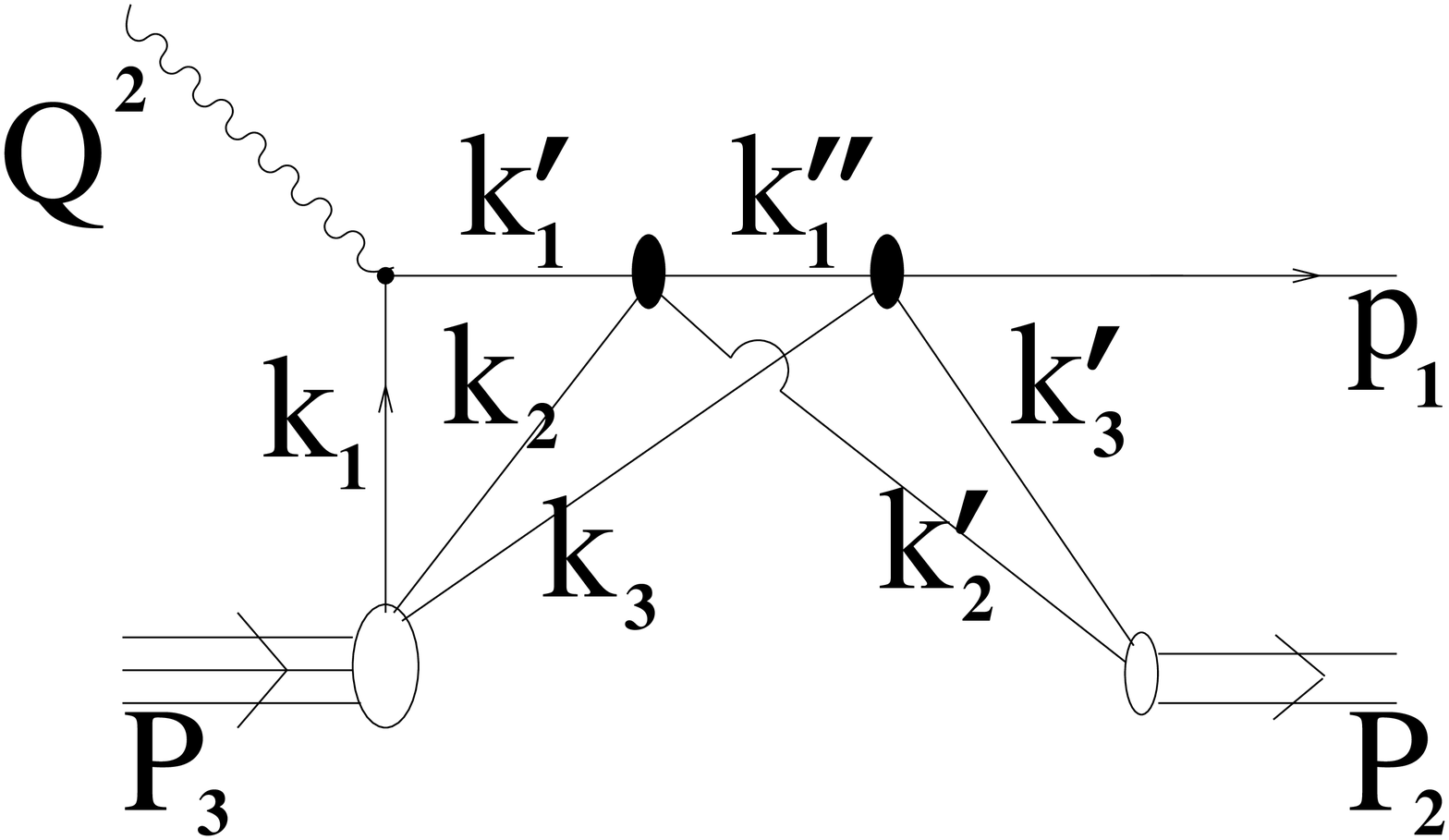}}
      \vspace{.5cm}
\centerline{\large\textbf{c)}\hspace{1cm}\normalsize}
\caption{The Feynman diagrams representing the PWIA (a)), the single (b)),
 and double (c)) rescattering in the
processes $^3He(e,e'p)D$ and $^3He(e,e'p)(np)$.
 In the former case the final two-nucleon state is a
deuteron with momentum ${\b P}_D={\b P}_2$,
whereas in the latter case the final state represents two
free nucleons with momenta ${\b p}_2$ and  ${\b p}_3$, with ${\b P}_2= {\b p}_2
+ {\b p}_3$.
 The trivial
single and double rescattering diagrams
with nucleons "2" and "3" interchanged are not drawn. The  black oval spots denote the elastic
nucleon-nucleon (NN) scattering
matrix ${\hat T}$ (See Eq. (25)).}
\label{fig6}
\end{figure}

\begin{figure}[!htp]
 \includegraphics[height=14cm,width=18cm]{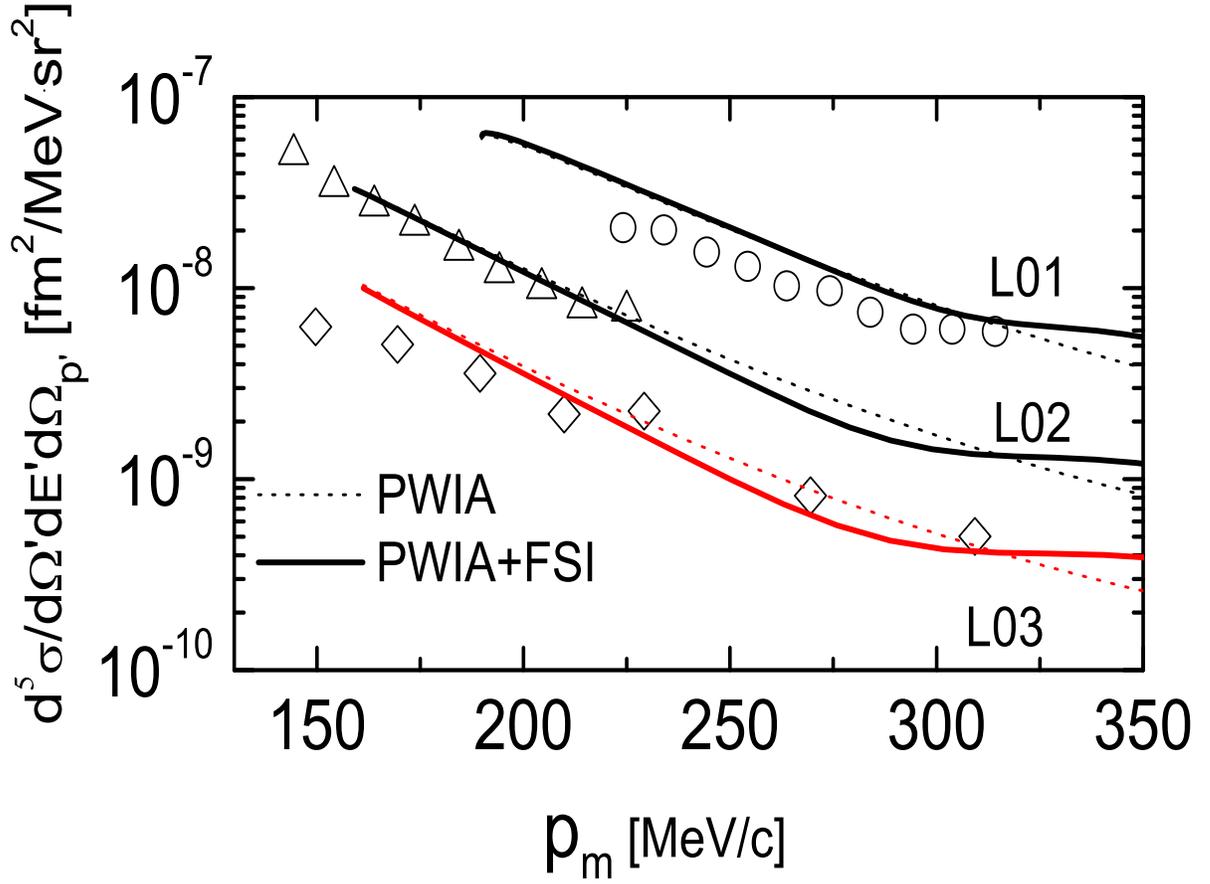}
\caption{The process $^2H(e,e'p)n$: the NIKHEF experimental data \protect\cite{nikhef}
{\it vs} the
missing momentum $p_m \equiv |{\b p}_m|$
are compared with our theoretical calculations; the dotted line represents the $PWIA$,
whereas the full line include the final state rescattering.
 The curves labelled $L01$,
$L02$, and $L03$,  correspond to $Q^2 = 0.1,\,\, 0.2,$ and
$0.3\,\,\, (GeV/c)^2$, respectively, and $x\simeq 0.3-0.6$ (in this Figure and in Figs. 8-12,
$p' \equiv |{\b p}_1|$).}
      \label{fig7}
\end{figure}

\begin{figure}[!htp]
 \includegraphics[height=10cm,width=11.0cm]{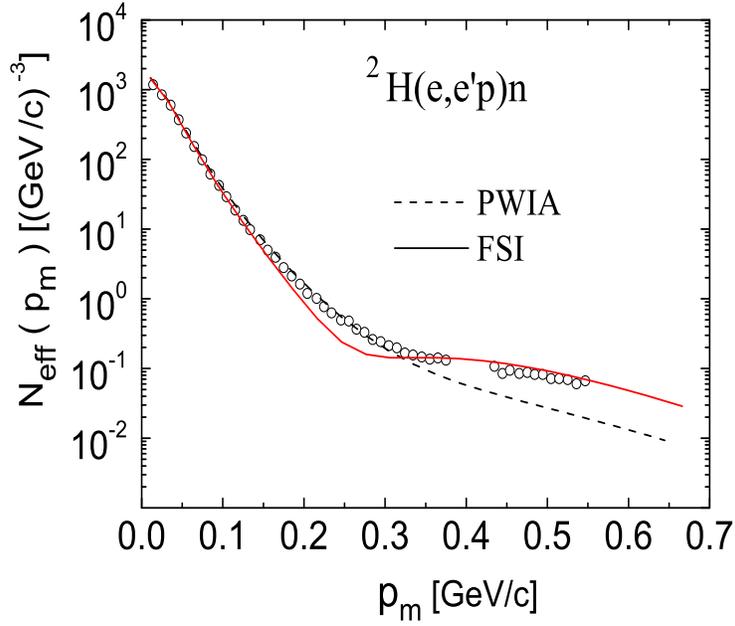}
\caption{The process $^2H(e,e'p)n$: the Jlab experimental data
 \protect\cite{ulmer}  ($N_{eff}$  defined by Eq.
(\ref
{neff}))
{\it vs} the
missing momentum $p_m \equiv |{\b p}_m|$,  compared with our theoretical calculations
.    The dotted line represents the PWIA
whereas the full line includes the final state rescattering. The experimental data correspond to the
perpendicular kinematics, with  $Q^2\simeq 0.665\,\,\, (GeV/c)^2$ , $|{\b q}|
\simeq 0.7\,\,\,
GeV/c$, and $x\simeq 0.96$.}
 \label{fig8}
\end{figure}

\begin{figure}[!htp]
 \includegraphics[height=12cm,width=13.0cm]{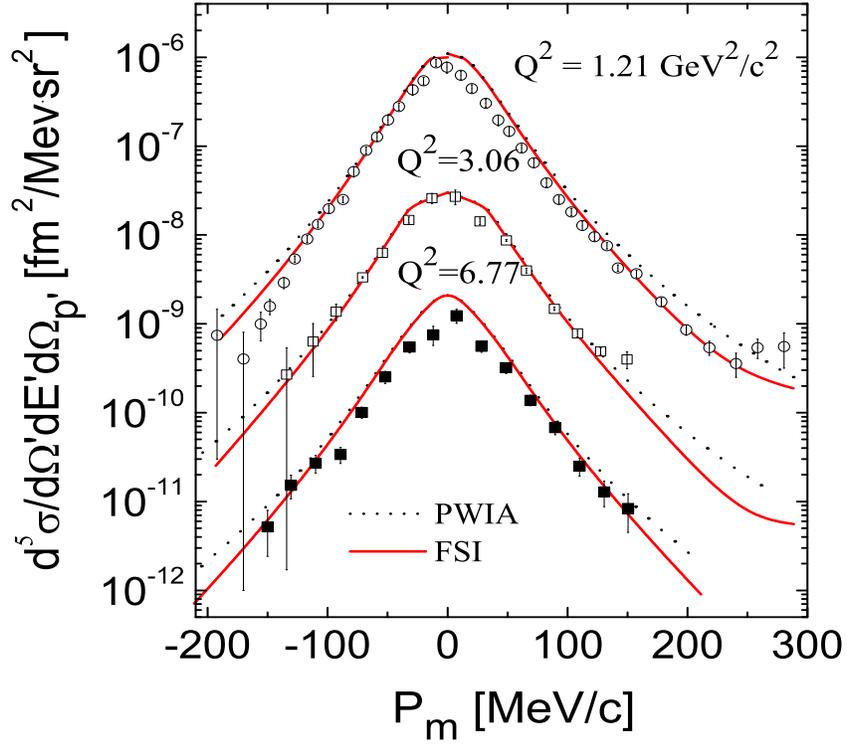}
      \caption{The process $^2H(e,e'p)n$: the SLAC  experimental data at $x\simeq 1$
 \protect\cite{bulten}
{\it vs} the
missing momentum $p_m \equiv |{\b p}_m|$,
compared with our theoretical calculations. The dotted lines represent
the PWIA
and  the full lines include the final state rescattering. The positive and negative
values of $p_m$ correspond to values of the azimuthal angle $\phi = \pi$ and $0$,
respectively.}
      \label{fig9}
\end{figure}

\begin{figure}[!hpt]
\includegraphics[height=14.0cm,width=14.0cm]{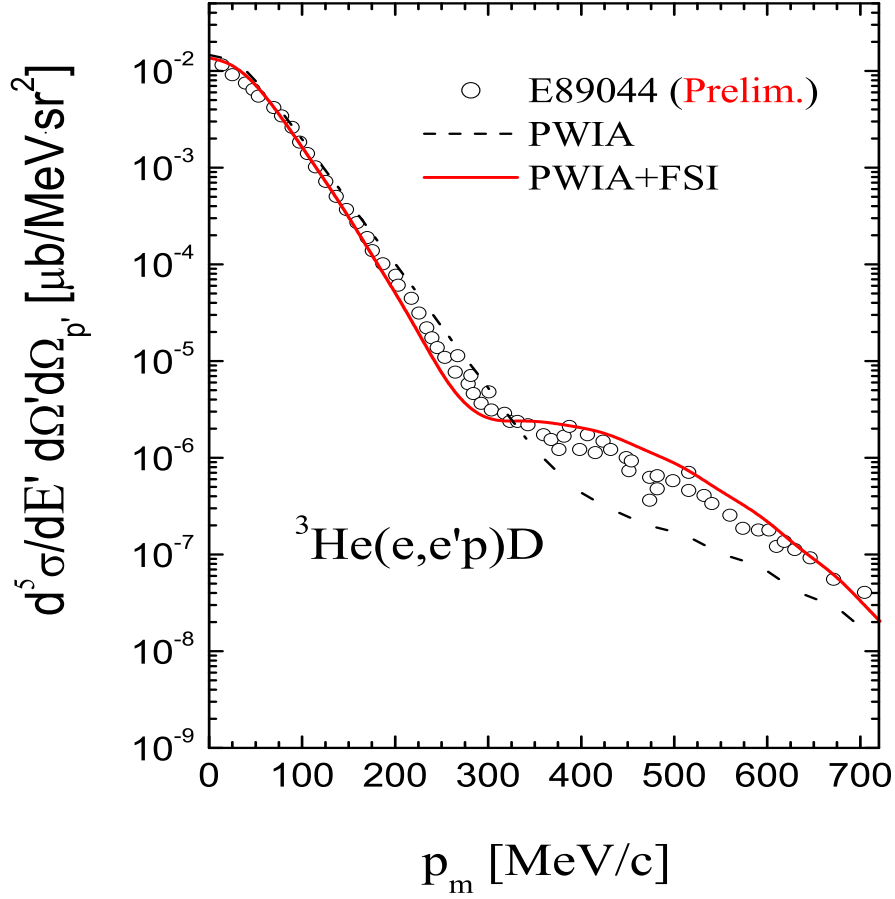}
\caption{The process $^3He(e,e'p)D$: the  experimental data from
JLab (JLab Experiment E-89-044 \cite{jlab1})
 {\it vs} $p_m \equiv |{\b p}_m|$   compared,  at
$Q^2 = 1.55\,\, (GeV/c)^2$ and $x=1$, with
our theoretical results. The dashed line corresponds to the
PWIA   and the full line
includes the full  FSI calculated using Eq. (\ref{ngrfsi}); the predictions by
 Eq.(\ref{totalS}) (GEA)  and Eq.  (\ref{eq20}) (GA),
  differ by at most $4\%$ and cannot be distinguished in the Figure
     (three-body wave function from \protect\cite{pisa},  $AV18$ interaction
     \protect\cite{av18}).}
 \label{fig10}
\end{figure}

\begin{figure}[!htp]
      \includegraphics[width=10cm,height=13.5cm]{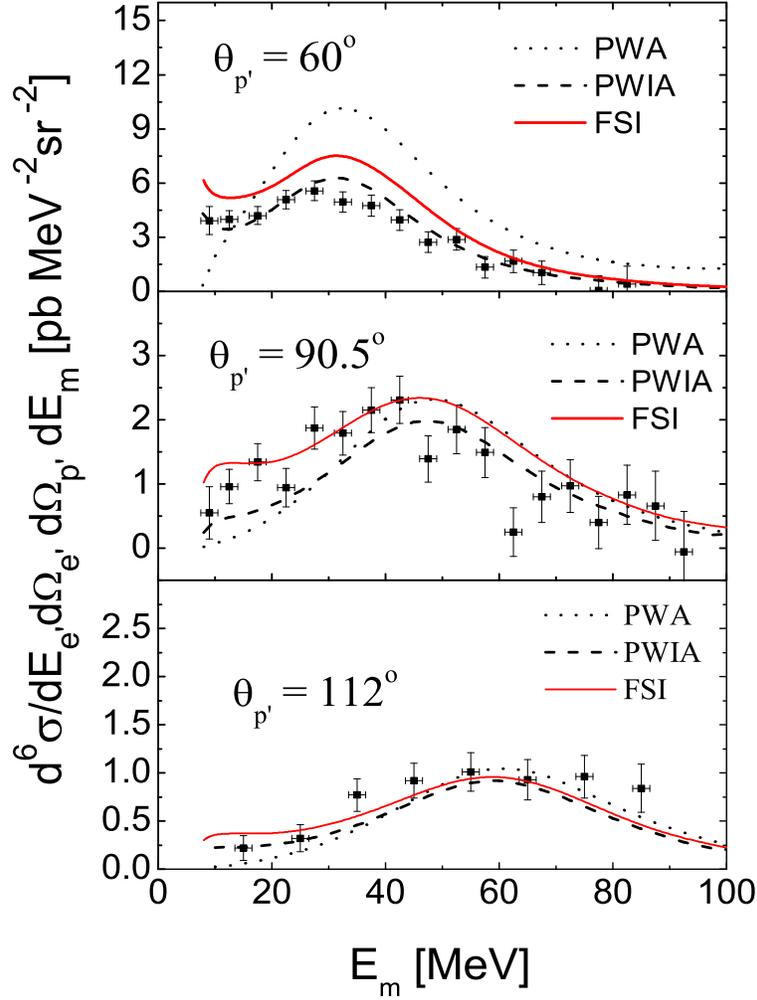}
\caption{The process $^3He(e,e'p)(np)$ : the experimental data from
Saclay \protect\cite{saclay} {\it vs} $E_m$ and for various values of the
proton emission angle $\theta_{{\b p}_1}$ ($p' \equiv |{\b p}_1|$), are   compared with our theoretical
results. The dotted lines correspond to the PWA,  when the three nucleons in the final
state are described by plane waves, the dashed lines correspond
 to the PWIA, when the interaction in the spectator neutron-proton pair
 is taken into account,  and the full lines
include the full  FSI calculated  using Eq. (\ref{ngrfsi}); the predictions by
 Eq.(\ref{totalS}) (GEA)  and Eq.  (\ref{eq20}) (GA),
  differ by at most $4\%$ and cannot be distinguished in the Figure.
Note that the values of the experimental $p_m$ and $E_m$ corresponding to  the maxima  of the cross section,
 satisfy to a large extent  the
relation  predicted by the
 two-nucleon correlation mechanism \cite{fscs}, namely $E_{m}\simeq
p_m^2/4M_N$  (cf. Fig. 3, right panel), with the  full FSI mainly affecting only the magnitude
of the cross section
     ( three-body wave function from \protect\cite{pisa},  $AV18$ interaction
     \protect\cite{av18}).}
\label{fig11}
\end{figure}

\begin{figure}[!hpt]
\includegraphics[height=14.0cm,width=8.0cm]{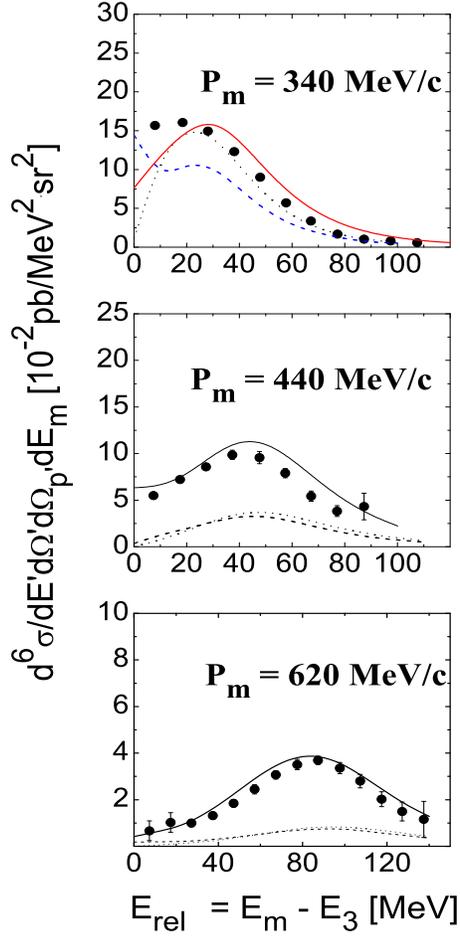}
\caption{  The process $^3He(e,e'p)(np)$: the same as in Fig.~\ref{fig11}
but for the Jlab experimental data  \cite{jlab2}) at $Q^2 = 1.55\,\, (GeV/c)^2$ and
$x=1$. Note, that unlike what shown in  Fig. \ref{fig11}, here
 the differential cross section is plotted {\it vs}
the excitation energy of the two-nucleon system in the continuum i.e.
  $E_{rel}=\displaystyle\frac{\b {t}^2}{M_N}$ = $E_2^f = E_m- E_{3}$.}
  \label{fig12}
\end{figure}

In Fig.~\ref{fig12}   our results are compared
  with the recent data from
Jlab \cite{jlab2}, where the cross section  was measured at fixed values
$|{\b p}_m|$ {\it vs.} the missing energy $E_m$.  As in the case of the Saclay data previously
analyzed, even in this case
  the cross section
 exhibits bumps approximately located
 at values of $E_m$ and $|{\b p}_m|$ satisfying the two-nucleon
 correlation mechanism
relation ($E_m\sim \displaystyle\frac{\b{p}_m^2}{4M_N}$),  and in
 agreement with the behaviour of the  Spectral Function
  (see Figs. \ref{fig3} and \ref{fig4}).
    However, contrary to the Saclay case, the PWIA dramatically
     underestimates the
    experimental data. This is clear evidence that the FSI
    between the struck nucleon and the nucleons of the
    spectator pair (Feynman diagrams
   b) and c)
   in Fig.~\ref{fig6}) does play a relevant role, as  the results of
    our calculations (the full line in Fig. \ref{fig12})  do indeed really show.
    Since, as already stressed, the Jlab experiment correspond to a perpendicular
    kinematics, this explains the  larger effects of the FSI  with respect
     to the Saclay
     experiment.
     The effects of the FSI calculated
 either within  the  GA  or GEA approximations,
 differ only by a few percent, which was
 expected in view of the observation
 that the factor ${\Delta_z}$ (Eqs. (\ref{delta3}) or (\ref{deltamark})) affects only
 the longitudinal component of ${\b p}_m$ and therefore has minor effects on
  the data we have  considered.
  
   There exist at present only two approaches to the calculations of the
   2bbu and 3bbu channels at the
    Jlab kinematics: the one presented in this paper and the one by Laget reported
    in Refs. \cite{jlab1}, \cite{jlab2} and \cite{laget5}.
    A comparison of the results of the two approaches exhibits an encouraging agreement
    both in the  2bbu and 3bbu channels, with some  minor differences which should
    most likely be ascribed
    to the different wave functions used in the two calculations. It is therefore
    gratifying to observe that
 different approximations to the treatment of FSI lead to  very similar  results.

    The effects of
     MEC and  $\Delta$, as previously pointed out, have not yet been
      considered in our approach;
    the calculation of  Ref. \cite{laget5}, shows they
  reduce the cross section in the peak by about $10 \%$,  leaving the
    missing energy dependence and, consequently, our conclusions,
   practically unchanged. 
    \section{Summary and Conclusions}
\label{sec:4}
 We have calculated the cross section of the
 processes  $^2H(e,e'p)n$, $^3He(e,e'p)D$, and  $^3He(e,e'p)(np)$, using realistic
 wave functions for the ground state, which exhibits  the very
 rich correlation structure
 generated by modern NN interactions;
 the FSI of the struck nucleon with the spectators has been treated
 within the standard Glauber eikonal  approximation (GA) \cite{glauber},
 as  well as with
   its generalized
 version (GEA) \cite{mark,strikman,misak}. The  two approaches differ
 by a factor ${\Delta_z}$ (Eqs. (\ref{delta}) and (\ref{deltamark}))
 which modifies
 (see Eq. (\ref{totalS})) the  FSI factor appearing in the standard GA
  (Eq. (\ref{eq20})).
 This factor takes into account in  the NN scattering amplitude the
 removal energy of the struck nucleon, or, equivalently, the excitation energy of the
 system $A-1$. By properly
 choosing the $z$-axis (along ${\b q}$ or
 ${\b p}_1$),  we were able to calculate   FSI  effects either
 in the case of large values of the three-momentum transfer
  ${\b q}$, or large values of the momentum of the struck nucleon
 ${\b p}_1$ relative to the $A-1$ system;
  by this way calculations could be extended
 successfully even at relatively
 low values of $Q^2$.
 As far as the  three-body break-up  channel in $^3He$ is concerned,
 the FSI in the spectator  pair was always calculated
 by the solution of the Schr\"odinger equation, whereas
  the interaction of the active, fast nucleon with the two nucleons of the
 spectator pair has been taken care of  by the GA or GEA approximations.
 The method we have used is a very transparent one and fully parameter free: it is based upon
 Eqs. (\ref{ddistr}), (\ref{pitotfsi}), and (\ref{totalS},
  which only require the knowledge of
 the nuclear wave functions, since the FSI factor is fixed directly
 by  NN scattering data. Of course
 with increasing $A$, the order of rescattering  increases up to the ($A-1$)-th order;
 we have performed calculations in the three-body case exactly,  and did not investigate
 the problem of the convergence
 of the multiple scattering series. This problem is under
 investigation in the case of $^4He$. Most of our calculations have been performed
 in kinematical conditions  where the effects of MEC, $\Delta$ isobar
 creation, etc. are  minimized, as confirmed by calculations
 performed, e.g., in Refs. \cite{laget1,laget2,laget3,laget4,laget5,hart1,hart2,hart3}.
   As for the main results we have obtained,
 the following remarks are in order:
 \begin{enumerate}
 \item the agreement between the results of our calculations and the
 experimental data for both the deuteron
 and $^3He$, is a very satisfactory one, particularly
  in view of the lack of any adjustable parameter
 in our approach;
 \item the effects of the FSI are such that they  systematically
  bring
  theoretical calculations in better agreement with the experimental data.
 For some quantities,  FSI simply improve the agreement
 between theory and experiment (cf.
 e.g. Figs. 7, 9 and 11), whereas for some other quantities,
 they play a dominant role (see
 e.g. Fig. 10 and 12);
\item a comparison of the PWA and the PWIA  with the full FSI calculation, does
 show that proper kinematics conditions could be found corresponding to an overall
 very small effect of FSI, leaving thus room for the investigation of the details
 of the nuclear wave function; as a matter of fact,
 we always found   that in the 3bbu channel in $^3He$, $^3He(e,e'p)(np)$,
  the experimental values of $p_m$ and $E_m$  corresponding  to the
   maximum values of the cross section,
satisfy to a large extent the relation  predicted by the
 two-nucleon correlation mechanism \cite{fscs}, namely $E_{m}\simeq
p_m^2/4M_N + E_3$  (cf. Fig. 3, right panel), with the  full FSI mainly
 affecting only the magnitude of the cross section;
 thus, quasi elastic one-nucleon emission  $A(e,e'p)B$ processes
 at $x\simeq 1$,
 together with processes at $x\simeq 2$,
 when the virtual photon is absorbed
 by a correlated two-nucleon "system",
 would represent valuable tool for the investigation of correlations in nuclei;
\item calculations of the 2bbu channel disintegration of  $^4He$, i.e.
the process $^4He(e,e'p)^3H$, have already been
 performed \cite{hiko} using realistic wave functions and
  taking exactly into account  nucleon rescattering up to 3rd order,
 i.e by using the generalization of Eq. (\ref{totalS}) to the four-particle case,
 {\it viz}
 \be
{\cal S}_{\Delta}^{FSI} ={\cal S}_{(1)}^{FSI}({\b R},\b{r}_{12},\b{r}_{34})+
{\cal S}_{(2)}^{FSI}({\b R},\b{r}_{12},\b{r}_{34})+ {\cal S}_{(3)}^{FSI}({\b R},\b{r}_{12},\b{r}_{34}),
\label{totalS4}
\ee
 where ${\b R}$, $\b{r}_{12}$, and $\b{r}_{34}$  are four-body Jacobi coordinates. Calculations
 for the 3bbu and 4bbu channels are in progress and will be reported elsewhere \cite{helium4}; they should in principle
 yield results appreciably differing from the predictions based upon shell-model type four-body wave functions;
 \item
 our results
 for $^3He$ generally agree with the ones obtained in Ref. \cite{laget5},
 so that it would appear that the problem of the treatment of FSI at high
 values of $Q^2$ (or high ${\bf p}_1$) is under control; nevertheless, a systematic comparison
 of the various approaches would be highly desirable;
 \item we have given the criteria according to which at high energies the exclusive
 $A(e,e'p)B$ cross section should factorize, and the similarity of our results withe the
 ones
 based upon a non factorized cross section \cite{laget5}, confirm the validity of these criteria.
 \item eventually, it appears that in the kinematical range we have considered
 only minor numerical differences were found between the conventional Glauber-eikonal
 approach and its generalized extension; this does not mean at all that the same
  will hold in other kinematical
 conditions (see e.g. \cite{strikman,mark,misak}).
 \end{enumerate}
 \section{Acknowledgments}
The authors are  indebted to  A. Kievsky  for making available
the variational three-body  wave functions of the Pisa Group and
to G. Salm\`e for useful discussions about their use.
Thanks are due  to M.A. Braun and D. Treleani for stimulating discussions on the
 Feynman diagram approach to nucleon rescattering.
Many useful discussions with various members of the Jlab
 Experiment E-89-044 are gratefully acknowledged.  We express our gratitude
  to M. Alvioli for a careful
 reading of the manuscript.
L.P.K. is   indebted to  the University of Perugia and INFN,
Sezione di Perugia, for a grant and for warm hospitality.
This work was partially supported by the Italian Ministero dell'Istruzione,
Universit\`{a} e Ricerca (MIUR), through the funds COFIN01,  and by the
 Russian Fund for Basic Research 00-15-96737.
\appendix
\section{ The nuclear wave functions}
\label{app}
In our calculations we have used two- and three-body  wave functions
corresponding to the  $AV18$ potential \cite{av18}.
\subsection{The ground state  wave function of  $^3He$ }
  For the $^3He$ wave function we have adopted the correlated
  variational wave function
  by the Pisa group \cite{pisa} which is  written in a
  mixed $(L_\rho,X,j_{23},S_{23})$-representation, where $j_{23}$ and $S_{23}$ are
  the total angular momentum and the total spin of the pair $"23"$, $X$ is an intermediate
  angular momentum resulting from the
  coupling  $\b {j}_{23} + \b {s}_1$ and $L_\rho$ is the
  radial angular momentum of the
  motion of the nucleon "1" relative to the pair "23". The explicit form of the wave
  function is
 \be &&
\Psi^{{\cal M}_{3}}_{He} (\mathbf{\boldrho,r}) = 
\sum\limits_{\{\alpha\}}\,\sum\limits_{\{m\}}
\la XM_X\ L_\rho m_\rho \ |\frac12 {\cal M}_3\ra
\la j_{23}m_{23}\ \frac12\sigma_1 \ | XM_X\ \ra
\chi_{\frac12 \sigma_1}\ {\rm Y}_{ L_\rho M_\rho}(\hat\boldrho)\nonumber\\&&
\la l_{23}\mu_{23}\ S_{23}\nu_{23} | j_{23}m_{23}\ra
 Y_{l_{23}\mu_{23}}(\hat\mathbf{r})\chi_{S_{23}\nu_{23}}
R_{\{\alpha\}}(r,\rho)\, {\cal I}_{\frac12\frac12}^{T_{23}},
\label{A1}
\ee
where $\{\alpha\}$ labels all  possible configurations in $^3He$ with
quantum numbers $L_\rho$,$X$,$j_{23}$,$S_{23}$, and $T_{23}$ and
$\la l_1  m_1 l_2 m_2   |\, l_{12} m_{12} \ra$ is a
Clebsch-Gordan coefficient. The total isospin function is
$ {\cal I}_{\frac12\frac12}^{T_{23}}=
 \sum\ \la \ T_{23}\tau_{23}\ \frac12\tau_1 \ | \frac12\frac12\ \ra
{\cal J}_{T_{23}\tau_{23}} \eta_{\frac12 \tau_1}$,
where ${\cal J}_{T_{23}\tau_{23}} $ and
$\eta_{\frac12 \tau_1}$ are  the
isospin functions of the pair and  the  nucleon, respectively.
Obviously, because of Pauli principle and  parity constraints,
the allowed configurations in eq. (\ref{A1}) are those that satisfy
the following conditions:
\be
L_\rho + l_{23} {\rm \,\,\,\,\,is\,\,\,\,
 even \,\,\,\,\,\,\,\,and \,\, \,\,\,\,\,\,l_{23}+S_{23}+T_{23} \rm\,\,\,\,\, is\,\,\, odd}
\label{A2}
\ee
The corresponding radial part of the wave function, $R_{\{\alpha\}}(r,\rho)$,
has been obtained \cite{pisa} by a variational method using  the $AV18$ potential
 including into the calculations values of $L_{\rho},l_{12}=0\ldots 9$
(a total of 58 different configurations  $L_\rho,X,j_{23},l_{23},S_{23}$
have been considered).
\subsection{The two-body continuum wave function $\Psi_{23}^{\b t}({\b r})$.}
With the  representation (\ref{A1})
 of the $^3He$ wave function, it was convenient to adopt for the  two-nucleon scattering state
 $\Psi_{23}^{\b t}({\b r})$   the spin-channel representation
 $\Psi_{S_{23}\nu_{23}}^{\bf t}({\bf r})$, characterized by the
 total (conserved in the scattering process) spin $S_{23}$ and its
 projection $\nu_{23}$. For  spin $S_{23}=1$ one has
 \be
 \Psi_{S_f\nu_f}^{\bf t}({\bf r})=
4\pi \sum\limits_{J_f M_f}\sum\limits_{l_0l_f}
\la l_0\mu_0\ S_f\nu_f | J_f M_f\ra
 {\rm Y}_{l_0\mu_0}(\hat{\bf t}) R_{J_f,l_0l_f}^{|{\bf t}|}(r)
 i^{l_f}{\cal Y}_{1l_f}^{J_f M_f}(\hat {\bf r}){\cal J}_{T_{23}\tau_{23}},
\label{A3}
 \ee
where $l_0,l_f=J_f\pm 1,\ J_f$. Note  that  the presence of tensor
forces in the NN-potential leads to an admixture of partial waves with
 $l=J_f-1$ and $l=J_f+1$. This hinders the use of real
 phase shifts for  the asymptotic behaviour  of the radial
 functions $R_{J_f,l_0l_f}^{|{\bf t}|}(r)$ and, consequently,
the Schr\"odinger equation cannot be solved in terms of real solutions.
 However, a unitary transformation $V$
 allows one  to define new radial functions $\tilde R=VR$
 which are  eigenfunctions of the scattering problem,
 i.e.,  solutions of the
 Schr\"odinger equation with the proper  asymptotic behaviour.
 \subsection{Wave function overlaps and the  spectral function $ P(|\b{k}_1|,E)$ of $^3He$.}
The Spectral function for the three-body break-up channel can be expressed in terms of the overlap between the three-body
and two-body radial functions by substituting  (\ref{A1})-(\ref{A3})
 into Eq. (\ref{piex}). Using the orthogonality of
 the spherical harmonics  ${\rm Y}_{lm}(\hat\b t)$
  and the completeness of the   Clebsch-Gordan
coefficients, one obtains that only  diagonal ($\{\alpha\}=\{\alpha_N\}$)
matrix elements contribute to the spectral function, viz.
  \be
 P_{ex}(|\b{k}_1|,E)&=&\frac12\sum\limits_{{\cal M}_{3}} \sum\limits_{\sigma_f,S_f,\nu_f}
 \int \frac{d^3 \b {t}}{(2\pi)^3}
  \left |
  \int d {\boldrho} d\b {r}
  \Psi_{He}^{{\cal M}_3}({\boldrho},\b{r})\Psi_{S_{f}\nu_{f}}^{\bf t}({\bf r})
  {\rm e}^{-i \boldrho \b {k}_1}
   \right |^2
\delta \left( E_m - \frac{\b {t}^2}{M_N} - E_3 \right)=\nonumber\\
&=&\frac{M_N\sqrt{M_NE_{rel}}}{2\pi^3} f_{iso}
\sum\limits_{\{\alpha\}} \left |
\int \rho^2 d\rho j_{L_\rho}(p\rho){\cal O}_{\{\alpha\}}^{E_{rel}}(\rho)
\right |^2,
\label{A4}
\ee
where $f_{iso}=3(1)$ for the pair in the isosinglet (isotriplet)
final state, $j_{L_\rho}(p\rho)$ is the spherical Bessel functions,  and
 the dimensionless overlap integrals ${\cal O}_{\{\alpha\}}^{E_{rel}}(\rho)$
are defined as follows
\be
{\cal O}_{\{\alpha\}}^{E_{rel}}(\rho)=
\int R_{\{\alpha\}}(r,\rho)\ \tilde
R_{\{\alpha\}}^{|{\bf t}|}(r) r^2 dr
\label{A5}
\ee
 The normalization of the proton spectral function (\ref{A4})-(\ref{A5}) is
 \be
 \int d^3{\b {k}_1} dE  P(|{\b k}_1|,E)\approx
 \left\{\begin{array}{cl}
 0.15& {\rm for } \ T_{23} = 0\\
  0.50& {\rm for }\  T_{23} = 1
 \end{array}
 \right.
 \label{A6}
 \ee
so that the
two-body break-up channel is normalized to  $\ \approx 1.35$.
 Since the FSI factors  $S^{FSI}$ and $S_{\Delta}^{FSI}$  (Eqs. (\ref{totalS}) and $(\ref{eq20})$)
are no spherically symmetric,
the  distorted
Spectral Function
$ P_{ex}^{FSI}({\b p}_m,E_m)$  (Eq.\ref{piexfsi})
 is not longer diagonal with respect
to the  $(L_\rho,X,j_{23},S_{23})$ configurations.
Except for parity constraints (\ref{A2}),  any values of
angular momenta of the pair in the final state
contribute to   $ P_{ex}^{FSI}({\b p}_m,E_m)$.
\section{Factorization of the covariant cross section}
\label{app:fact}
In this Appendix we will show, within a fully covariant approach, that under certain
kinematical conditions the cross section for the process $A(e,e'p)X$ process
factorizes even in presence of FSI. We shall consider,
to this end, the deuteron (D) treated within the Bethe-Salpeter (BS) formalism.
As mentioned, the factorization depends upon the
spin structure of the square of the matrix element
$
\left [ \bar u (\b{k}_1,\tilde{s_1}) \Phi_{D}^{{\cal M}_2}(k_1,k_2)
 (\hat k_2+M_N) v(\b{k}_2,s_2)\right ]
$ appearing in Eq.  (\ref{ampli})
or, in case of FSI, upon the structure  of
$ \left [ \bar u(\b{k}_1,\tilde{s_1})
 \Phi_{D}^{{\cal M}_2}(k_1,k_2) v(\b{k}_2,{s_2})\right]
$ (cf Eq. (\ref{ttot})).
The relevant spin parts can be evaluated directly by
using the explicit form of the Dirac spinors, $u$ and $v$, and
the explicit expressions for the amplitudes
$\Phi_{D}^{{\cal M}_2}(k_1,k_2)$ (cf. Refs.~\cite{deuteronrel,gross,amplBS}).

\subsection{The PWIA}
In Ref. \cite{bs} the Feynman diagrams
for the process  $D(e,e'p)X$ have been evaluated
including all BS components. Here
we re-calculate the diagrams for the $\Spp$ and $\Dpp$  components
in a slightly different manner which will be useful when FSI effects are considered.

In PWIA the cross section reads as follows
\be
\frac{d^5\sigma}{d E' d\Omega'}=
\sigma_{Mott}\tilde l^{\mu\nu}L_{\mu\nu}^D\frac{d^3p_1}{(2\pi)^3 2E_1}
\frac{d^3p_2}{(2\pi)^3 2E_2},
\label{cross}
\ee
where  $\tilde l^{\mu\nu}$  and  $L_{\mu\nu}^D$  are the leptonic and hadronic tensors,
respectively, the latter being
\be &&
L_{\mu\nu}^D=\frac{1}{2M_D}\frac13\sum\limits_{{{\cal M}_2},s_1,s_2}
T_\mu ({{\cal M}_2},s_1,s_2) T_\nu ({{\cal M}_2},s_1,s_2)
(2\pi)^3\delta^{(4)}\left( P_D+q-p_1-p_2\right)=\label{tensor}\\
&&=\frac{1}{2M_D}\frac13\sum\limits_{{{\cal M}_2},s_1,s_2}
\la {{\cal M}_2}|{\hat J_{\mu}^N} |p_2,s_2,p_1,s_1\ra
\la p_1,s_1,p_2,s_2 |{\hat J_{\nu}^N} |{{\cal M}_2}\ra
(2\pi)^3\delta^{(4)}\left( P_D+q-p_1-p_2\right),
\nonumber
\ee
where
${\hat J_{\mu}^N}$ is the nucleon electromagnetic current operator.
The amplitude $T_\mu$ could be  written in the following form
\be
T_\mu ({{\cal M}_2},s_1,s_2)=\bar u(\b {p}_1,s_1)
\Gamma_\mu^{\gamma^* N} (Q^2,k_1^2) \Phi_D^{{{\cal M}_2}}(k_1,k_2)
\tilde S^{-1}(\hat k_2) v(\b{p}_2,s_2)
\label{amplit},
\ee
where $\Phi_D^{{\cal M}_2}$ is a short-hand notation for the
main BS amplitudes  $\Phi_{\Spp}$ and
$\Phi_{\Dpp}$, corresponding to  $L=0$ and $L=2$, respectively
(see Refs. \cite{deuteronrel,amplBS}),
$k_2=p_2$,
 ${\hat S}^{-1}(\hat k_2)=\hat k_2 + m$,
 and $\Gamma_\mu^{\gamma^*N} (Q^2,k_1^2) $
is the electromagnetic $eN$ vertex  which, for an off-mass-shell nucleon,
depends not only upon $Q^2$, but upon $k_1^2\neq m^2$ as well.

By introducing  between $\Gamma_\mu (Q^2,p_1,k_1)$ and $ \Phi_D^{\CM_D}(k_1,k_2)$ the
complete set of the Dirac spinors
\be
\dfrac{1}{2M_N}
\sum\limits_{\tilde s_1}\left[ u(\b{k}_1,\tilde s_1) \bar u(\b{k}_1,\tilde s_1)-
v(\b{k}_1,\tilde s_1)\bar v (\b{k}_1,\tilde s_1)\right]
\label{sp}
\ee
and bearing in mind that for the $\Spp$ and $\Dpp$
partial waves the second term in (\ref{sp}) does not contribute, we obtain
\be
T_\mu ({\cal M}_2,s_1,s_2)= \dfrac{1}{2M_N}
\sum\limits_{\tilde s_1} J_\mu^{eN}(Q^2,p_1,k_1,\tilde s_1,s_1) 
 \left [ \bar u (\b{k}_1,\tilde s_1) \Phi_D^{M}(k_1,k_2)
\tilde S^{-1}(\hat k_2) v(\b{p}_2,s_2)\right ],\,\,
\label{amplit1}
\ee
where
$ J_\mu^{eN}(Q^2,p_1,k_1)=\la \b{p}_1,s_1 | \Gamma_\mu^{\gamma^*N}
(Q^2,k_1^2) | \b{k}_1,\tilde s_1\ra$.

Let us evaluate Eq. ({\ref{amplit1}}) for  the $D$ wave. One has
\be
&&\left [ \bar u (\b{k}_1,\tilde s_1) \Phi_{\Dpp}^{{\cal M}_2}(k_1,k_2)
\tilde S^{-1}(\hat k_2) v(\b{p}_2,s_2)\right ]= \nonumber\\
&=&-\dfrac{N N_1^2}{\sqrt{2}}
(k_2^2-M_N^2)\phi_D(k_0,|\b{k}|)\ 2m
\left\langle
\chi_{\tilde s_1}^{\dagger}\left | \left \{-(\boldsigma\boldxi^M)+
3(\b{n}\boldxi^M)(\b{n}\boldsigma)\right\}\right |
\widetilde\chi_{s_2}\right\rangle,
\label{eq3}
\ee
where $\b n$ is a unit vector along $\b k$, i.e.,
$\b {n}=\dfrac{\b{k}}{|\b {k}|}=\dfrac{\b{k}_1}{|\b {k}_1|}$.
When Eq. (\ref{eq3}) is inserted in the expression for the cross section,  one obtains

\begin{eqnarray}
&&
\dfrac13 \sum\limits_{{{\cal M}_2},s_2}
\left\langle
\chi_{\tilde s_1}\left | \left \{-(\boldsigma\boldxi^{{\cal M}_2})+
3(\b{n}\boldxi^{{\cal M}_2})(\b{n}\boldsigma)\right\}\right
|\widetilde\chi_{s_2}\right\rangle
\left\langle
\widetilde\chi_{s_2}\left | \left \{-(\boldsigma\boldxi^{+{{\cal M}_2}})+
3(\b{n}\boldxi{^+M})(\b{n}\boldsigma)\right\}\right
|\chi_{\tilde s_1}\right\rangle
\nonumber\\ &&
=2\delta_{\tilde s_1{\tilde s_1}'}.
\label{eq4B}
\end{eqnarray}
The last relation ensures factorization of the cross section; as a matter of fact,
by performing the same procedure for the $S$-wave, it easy to show that, thanks to
Eq. (\ref{eq4B}),
 the cross section (Eq. (\ref{cross})) factorizes, assuming the form
 (\ref{eq19}) with  $n_D$ given by Eq. (\ref{momdis}). In obtaining the
 above equations we expressed the
BS amplitudes $\phi_L(k_0,|\b{k}|)$ in terms of the BS
vertices   ${G_{{^3{L^{++}_1}}}(k_0,|\b{k}|)}$ and the radial functions
$u_L$,  by the relations
\be
\dfrac{NN_1^2}{\sqrt{2}}(k_2^2-M_N^2)\phi_D(k_0,|\b{k}|)=
\dfrac{NN_1^22E_{\b k}}{\sqrt{2}}\dfrac{G_{\Dpp}(k_0,|\b{k}|)}{M_D-2E_{\b k}},
\ee
where $k_0=\dfrac{M_D}{2}-E_{\b k}$, and
\be
u_{S(D)}=\dfrac{G_{\Spp (\Dpp)}(k_0,|\b{k}|)/(4\pi)}{\sqrt{2M_D}(M_D-2E_{\b k})}.
\label{waves}
\ee
Note that in Eq. (\ref{waves}) the normalization of the wave function is chosen so
as
to correspond to the non relativistic deuteron wave function
\be
\dfrac2\pi \int |\b{k}|^2 d  |\b{k}| \left(
u_S^2 (|\b{k}|) + u_D^2 (|\b{k}|) \right)\approx 1.
\ee
We reiterate that factorization in PWIA occurs because
 the sum over $s_2$ and ${{\cal M}_2}$
of the square of the matrix element in Eq. (\ref{eq4B} ) becomes  diagonal
with respect to $\tilde s_1$.

When FSI is taken into account, instead of Eq. (\ref{eq4B}), one
obtains for the $D$-wave
(for the $S-$wave the spin structure is
trivial)
 \be &&
 \frac13\sum\limits_{{\cal M}_2,s_2}
  \left [ \bar u(\b{k}_1,{s_1})
 \Phi_{D}^{{\cal M}_2}(k_1,k_2) v(\b{k}_2,{s_2})\right]^\dagger
  \left [ \bar u(\b{k}'_1,\tilde{s_1})
 \Phi_{D}^{{\cal M}_2}(k_1',k_2') v(\b{k}'_2,{s_2})\right]
 \simeq\nonumber\\&&
 \dfrac13 \sum\limits_{M,s_2}
\left\langle
\chi_{s_1}\left | \left \{-(\boldsigma\boldxi^{{\cal M}_2})+
3(\b{n}\boldxi^{{\cal M}_2})(\b{n}\boldsigma)\right\}\right
 |\widetilde\chi_{s_2}\right\rangle
\left\langle
\widetilde\chi_{s_2}\left | \left \{-(\boldsigma\boldxi^{+{{\cal M}_2}})+
3(\b{n}'\boldxi^{+{{\cal M}_2}})(\b{n}'\boldsigma)\right\}\right
|\chi_{\tilde s_1}\right\rangle=\nonumber\\&&
=\dfrac13 \sum\limits_{{{\cal M}_2}}
\left\langle
\chi_{s_1}\left |
\left \{
-(\boldsigma\boldxi^{{\cal M}_2})+ 3(\b{n}\boldxi^{{\cal M}_2})(\b{n}\boldsigma)
\right\}
 \left \{
 -(\boldsigma\boldxi^{+{{\cal M}_2}})+ 3(\b{n}'\boldxi^{+{{\cal M}_2}})(\b{n}'\boldsigma)\right\}
 \right |
 \chi_{\tilde s_1}
 \right \rangle
 \label{eq44app}
\ee
 where $\bf n$  (${\bf n}'$)
 is a unit vector along ${\bf k}_1$ (${\bf k}'_2$).

 By taking into account the  completeness of the
 polarization vectors $\boldxi$, the only
  spin dependence remaining  in Eq. (\ref{eq44app}) is contained in  the term
 \be &&
 \frac13 \sum_M
 (\b{n}\boldxi^M)(\b{n}\boldsigma)(\b{n}'\boldxi^{+M})(\b{n}'\boldsigma)=
 (\b{n}\b{n}')\left(\phantom{\!\!\!\frac11}
 (\b{n}\b{n}') -i\boldsigma\left[\b{n} \times \b{n}'\right]\right),
 \label{ads}
 \ee
 so that in case of rescattering with low momentum transfer, when in  the integral
 over $\b{k}_2$ the main contribution comes from
  $\b{k}_2\sim\b{p}_2\,\, , \b{k}_2'\sim\b{p}_2$,
  one has $\boldsigma\left[\b{n} \times \b{n}'\right]=0$
 and  factorization is approximately recovered,  with
 the $S$ and $D$ waves adding  incoherently.

  Thus, to sum up,  factorization is compatible  with FSI if:
  \begin{enumerate}

\item The spin-flip part of the $NN$ amplitude should be very small, as it occurs
when  either the three-momentum transfer $\b q$,\, or
 the momentum  $|{\b p}_1|$  are large;

\item the momentum transfer $\boldkappa$ in the $NN$ rescattering has to be small
small so that in the integral
    $\b{k}_2\sim\b{p}_2$. This appears to be the case since
     the $NN$
    amplitude  is sharply  peaked forward.

\item The contribution from  $N\bar N$ pair currents can be neglected,
which is to a large extent legitimate due to the  smallness of the $P$ wave in the deuteron.
\end{enumerate}

\newpage
\vskip 2mm
\newpage
\vskip 2mm

 \newpage
\newpage
\newpage

\newpage

\newpage

\newpage

\newpage

\newpage

\newpage

\newpage

\end{document}